\documentclass[]{raa}            % referee version: for submission
\usepackage{graphicx,times}             %for PS/EPS graphics inclusion, new
\usepackage{natbib}
\usepackage{amssymb,amsmath}
\usepackage[OT1]{fontenc} 
\usepackage{diagbox}
\usepackage{multirow}

\bibpunct{(}{)}{;}{a}{}{,}

\usepackage[a4paper=true,pagebackref=true]{hyperref}%a4paper=true,pagebackref=true
\hypersetup{colorlinks = true, linkcolor = green, anchorcolor = red, citecolor = blue, filecolor = red, pagecolor = red, urlcolor = red}
\begin{document}

   \title{What causes the absence of pulsations in Central Compact Objects in Supernova Remnants?
%\footnotetext{$*$ Supported by the National Natural Science Foundation of China.}
}
%   \subtitle{I. Place Your Subtitle Here}

   \volnopage{Vol.0 (20xx) No.0, 000--000}      %%preserved for Editor. DOn't remove!
   \setcounter{page}{1}          %%starting page, preserved for Editor. DOn't remove!

   \author{Qi Wu
      \inst{1,2}
   \and Adriana M. Pires
      \inst{1,3}
   \and Axel Schwope
      \inst{3}
   \and Guang-Cheng Xiao
   	\inst{1}
   \and Shu-Ping Yan
   	\inst{1}
   \and Li Ji
      \inst{1,2}
   }

   \institute{Purple Mountain Observatory, Key Laboratory of Dark Matter and Space Astronomy, Chinese Academy of Sciences, Nanjing 210033, People's Republic of China; {\it wuqi@pmo.ac.cn}\\
        \and
             School of Astronomy and Space Science, University of Science and Technology of China, Hefei 230026, People's Republic of China\\
        \and
             Leibniz-Institut f\"{u}r Astrophysik Potsdam (AIP), An der Sternwarte 16, 14482 Potsdam, Germany; {\it apires@aip.de}\\
\vs\no
   {\small Received~~2021 month day; accepted~~2021~~month day}}

\abstract{ Most young neutron stars belonging to the class of Central Compact Objects in supernova remnants (CCOs) do not have known periodicities. We investigated seven such CCOs to understand the common reasons for the absence of detected pulsations. Making use of XMM-Newton, Chandra, and NICER observations, we perform a systematic timing and spectral analysis to derive updated sensitivity limits for both periodic signals and multi-temperature spectral components that could be associated with radiation from hotspots on the neutron star surface. Based on these limits, we then investigated for each target the allowed viewing geometry that could explain the lack of pulsations. We estimate it is unlikely ($<10^{-6}$) to attribute that we do not see pulsations to an unfavorable viewing geometry for five considered sources. Alternatively, the carbon atmosphere model, which assumes homogeneous temperature distribution on the surface, describes the spectra equally well and provides a reasonable interpretation for the absence of detected periodicities within current limits. The unusual properties of CCOs with respect to other young neutron stars could suggest a different evolutionary path, as that proposed for sources experiencing episodes of significant fallback accretion after the supernova event.
\keywords{methods: data analysis --- stars: neutron --- X-rays: stars}
}

   \authorrunning{Q. Wu et al. }            %author_head in even pages
   \titlerunning{What causes the absence of pulsations in the CCOs?}  % title_head in odd pages

   \maketitle
%% The author head (on even pages) and the title head (on odd pages) will be
%% automatically extracted from \author{} and \title{}. Whenever the title is too long,
%% you will be asked to supply a shorter one by inserting either \authorrunning{} or
%% \titlerunning{} before \maketitle. Anyway, you can specify your own heads.
%%
%%
%% Note: In the following text body of your manuscript, please note several differences from
%%       other major journals:
%% (1) \subsection{Please Capitalize the First Letter of Each Notional Word in Subsection Title}
%% (2) Please Capitalize the First Letter of Each Notional Word in all tables' captions

%
%_________________________ sections below _______________________ 
%
\section{Introduction}
\label{sect:intro}

\par The Central Compact Objects (CCOs) are a handful of young neutron stars, which are found close to the center of young ($0.3-7\,$kyr) supernova remnants (SNRs) (see \citealt{deluca2008AIPC..983..311D,deluca2017JPhCS}, for reviews). The discovery of these objects is limited by observational technology. CCOs share similar properties, such as the lack of known optical/radio/IR counterparts; their purely thermal-like spectra can usually be described by a two-blackbody model with a temperature of $0.2-0.5\,$keV (\citealt{deluca2017JPhCS}). At present, about ten sources are confirmed members of the class\footnote{See the complete catalog at \url{http://www.iasf-milano.inaf.it/~deluca/cco/main.htm}}. Only for three sources have periodicities been detected in the X-ray band ($105-424\,$ms; \citealt{Zavlin2000ApJ25Z,gotthelf2005ApJ390G,gotthelf2009ApJ...695L..35G}).

\par Dedicated campaigns with XMM-Newton and Chandra revealed the very low period derivatives of these three CCOs ($8.7\times10^{-18}-2.2\times10^{-17}\,\mathrm{s\,s^{-1}}$; \citealt{Halpern2010ApJ436H,Gotthelf2013ApJ58G,halpern2015ApJ61H}). Under the assumption of the standard magnetic dipole braking model (\citealt{ostriker1969ApJ1395}), their characteristic ages are within $192-301\,$Myr (\citealt{Halpern2010ApJ436H,Gotthelf2013ApJ58G,halpern2015ApJ61H}), which implies a factor of about five order magnitudes higher than the estimated age of the associated SNRs ($4.5-7\,$kyrs; \citealt{Sun2004ApJ742S,Becker2012ApJ141B,Roger1988ApJ940R}). Even though the magnetic field strength of the three CCOs are within $2.9-9.8\times10^{10}\,$G (\citealt{Gotthelf2013ApJ58G,Halpern2010ApJ436H,halpern2015ApJ61H}), the anisotropic temperature distributions on the CCOs could be due to localized crustal heating by magnetic decay, which requires a magnetic field strength higher than $10^{14}\,$G inside the crustal region (\citealt{Halpern2010ApJ436H}). Alternatively, it may be caused by anisotropic heat transfer from the interior, which also points to the possibility of stronger components of the magnetic field inside the neutron star (\citealt{Perez2006AA1009P,geppert2004AA267G}). 
A normal magnetic field may be buried inside the core or crust of the neutron star so that a weak magnetic field presents. The buried magnetic field then takes thousands of years to diffuse back to the surface, during which time the neutron star shows a weak magnetic field and with no flux variability, which is so-called ``anti-magnetar'' (\citealt{Halpern2010ApJ436H,Gotthelf2013ApJ58G}). Though the most CCOs have low pulsed fractions\footnote{PF=($F_{max}-F_{min}$)/($F_{max}+F_{min}$), where $F_{max}$ and $F_{min}$ are the maximum and the minimum fluxes} (see, e.g., \citealt{deluca2008AIPC..983..311D,halpern2010ApJ941H}), the ``anti-magnetar'' explanation still can not account for the high pulsed fraction (hereafter, PF) that occurred in Kes 79 (\citealt{Halpern2010ApJ436H}) or the two antipodal hot spots of different temperatures and areas on the CCO in Puppis A (\citealt{gotthelf2009ApJ...695L..35G}) since the weak magnetic fields are not enough to generate the luminosity in X-ray band (\citealt{Gotthelf2013ApJ58G}).

\par Alternatively, the carbon atmosphere model, which relies on a uniform temperature distribution on the surface, successfully described the spectrum of the CCO in Cas A and also yielded a neutron star radius expected for nuclear matter (\citealt{Ho2009Natur71H,ozel2016ARAA..54..401O}). This model was also succesfully applied to the other CCOs, such as XMMU J173203.3-344518 in G353.6-0.7 (\citealt{klochkov2013AA}) and CXOU J160103.1-513353 in G330.2+10 (\citealt{doroshenko2018AA76D}). Nevertheless, there is no direct evidence to verify the existence of a carbon atmosphere (\citealt{Bogdanov2014ApJ94B,suleimanov2017AA43S}). Notably, \cite{Bogdanov2014ApJ94B} showed that the spectrum of the CCO in Kes 79 can also be well described by a single-temperature carbon atmosphere model despite the high pulsed fraction of the source ($64\%$; \citealt{Halpern2010ApJ436H}).

\par On the other hand, an unfavorable viewing geometry can also naturally explain the absence of pulsed signals (\citealt{suleimanov2017AA43S,doroshenko2018AA76D,Pires2019AA73P}). Although definite conclusions in general cannot be drawn for individual sources, it is possible to estimate the joint probability that we miss modulations for all neutron stars belonging to this class \citep[e.g.][]{doroshenko2018AA76D}. By modeling light curves of three objects, \citeauthor{doroshenko2018AA76D} estimated a small joint probability for such a scenario, about 0.3\%.

\par In this work, we analyze seven CCOs without detected periods as a class to investigate the common reasons for the absence of pulsations. We consider the observations operated by XMM-Newton, Chandra, and NICER for more stringent limitations on the pulsed fraction. This paper is organized as follows: In Section~\ref{sect:Obs}, we describe the observations and data reduction of XMM-Newton, Chandra, and NICER. We presented the data analysis and the results in Section~\ref{sect:analysis}. In Section~\ref{sect:discussion}, we apply our results to discuss the viewing geometry of the CCOs and the presence of hot spots. We summarize the results and conclusions in Section~\ref{sect:conclusion}.

\section{Observations and Data reduction}
\label{sect:Obs}

\par All targets have been repeatedly observed by XMM-Newton (\citealt{Jansen2001AA1J}), Chandra Observatories (\citealt{Weisskopf2002PASP1W}) and the Neutron star Interior Composition Explorer (NICER, \citealt{Gendreau2017NatAs895G}) in several occasions in the past twenty years. Table~\ref{tab:obs_gti} gives an overview of the datasets included in the analysis, with details of the scientific exposures and corresponding observing modes. For each target we analysed the longest archival observation with as high time resolution as possible, to achieve the most sensitive and constraining limits for the analysis. For Cas A, we did not include the observation operated by XMM-Newton because of heavy contamination from the supernova remnants ($\sim90\%$), even though it was observed longer and with higher time resolution than before ($117\,$ks; ObsID 0650450201).
The observations conducted with the EPIC-pn camera (\citealt{Struder2001AA18S}) in small-window (SW) and full-frame (FF) imaging mode provide a time resolution of 5.7\,ms and 73.4\,ms, respectively. In contrast, the time resolution of the EPIC-MOS can only reach 0.3 s in SW mode and 2.6 s in FF mode. The Advanced Camera for Imaging and Spectroscopy (ACIS-S, \citealt{Garmire2003SPIE28G}) in FAINT and continuous-clocking (CC) mode provide time resolutions of 0.341 s and 2.85 ms, respectively. Moreover, the X-ray Timing Instrument (XTI), NICER's payload, provides time resolution better than $100\,$ns (\citealt{Gendreau2017NatAs895G}).

\par For the NICER observations, we estimated the level of background contamination using $\texttt{PIMMS}$\footnote{\url{https://heasarc.gsfc.nasa.gov/cgi-bin/Tools/w3pimms/w3pimms.pl}} and the spectral parameters of each CCO determined in previous work. For the CXOU J232327.9+584842 in Cas A, the photon count is dominated by the emission from the supernova remnant (over 99\%), so we do not include it in the analysis. Furthermore, we found that the observation of the CXOU J181852.0-150213 provides more photons so that it is possible to get a stringent PF limit than before. Besides, we consider the observation of 1WGA J1713.4-3949 and XMMU J173203.3-344518 here.

\begin{table}[!t]
	\centering\small
	\caption{Observations of the CCOs}
	\label{tab:obs_gti}
	\begin{tabular}{ccccccccr}
		\hline
			Target & Target ID & Obs. ID & Date & Inst. & Mode &  Duration  & GTI & Net Counts$^{a}$\\ 
			& & & & & & (ks) & (\%) & \\
		\hline
         CXOU J085201.4-461753 & J0852 & 0652510101 & 2010-11-13 & pn   & SW & 84  & 78.4 & $2.56(16)\times10^4$ \\
         CXOU J160103.1-513353 & J1601 & 0742050101 & 2015-03-08 & pn   & FF & 138 & 63.7 & $2.28(5)\times10^3$ \\
        	1WGA J1713.4-3949 & J1713 & 722190101 & 2013-08-24 & pn   & SW & 137 & 92.5 & $1.174(3)\times10^5$ \\
            &  & 3201030101 & 2020-03-11 & XTI & PHOTON & 11.5 & 100 & $3.29(2)\times10^4$ \\
         XMMU J172054.5-372652 & J1720 & 14806     & 2013-05-11 & ACIS-S & CC & 89 & 99.9 & $4.22(6)\times10^3$ \\
        	XMMU J173203.3-344518 & J1732 & 0722190201 & 2014-02-24 & pn   & SW & 128 & 86.9 & $5.20(23)\times10^4$\\
				 & & 0722090101 & 2013-10-05 & pn   & FF & 58  & 90.9 & $2.26(15)\times10^4$\\
        		&  & 0694030101 & 2013-03-07 & pn   & FF & 69  & 78.7 & $2.42(15)\times10^4$\\
            &  & 1030230107 & 2018-03-23 & XTI  & PHOTON   & 16  & 100 & $2.057(14)\times10^4$ \\
        	CXOU J181852.0-150213 & J1818 &16766	   & 2015-07-30 & ACIS-S & VFAINT & 92 & 99.6 & $275\pm16$ \\
            &  & 0034130102 & 2017-07-14 & XTI & PHOTON & 9 & 100 & $2.37(5)\times10^3$ \\
         CXOU J232327.9+584842 & J2323 &16946      & 2015-04-27 & ACIS-S & FAINT & 68 & 100 & $5.58(7)\times10^3$ \\
		\hline
	\end{tabular}
	\note{The EPIC cameras were operated in imaging mode, while the ACIS-S were operated in imaging mode (FAINT and VFAINT) and timing mode (CC). We list the percentage of good-time-intervals (GTIs) after filtering out periods with high background flare. The GTIs of those Chandra observations are close to 100\% since it has been filtered out by using the task \texttt{chandra\_repro}. $^{(a)}$ The net counts were calculated at the energy range of $0.5-8.0\,$keV. }
\end{table}

\par For XMM-Newton observations, we performed standard data reduction with SAS version 18.0.0 (xmmsas20190531\_1155-18.0.0) applying the latest calibration files and following the analysis guidelines of the EPIC-pn instrument\footnote{\url{https://xmmweb.esac.esa.int/docs/documents/CAL-TN-0018.pdf}}. We processed the exposures using the SAS meta task $\texttt{epproc}$ and applied default corrections. For the Chandra observations, we used $\texttt{ciao}$ 4.12 (Chandra Interactive Analysis of Observations) with the corresponding CALDB version 4.9.2.1 for the data processing. For NICER, standard data reduction was performed with HEASoft v.6.28, using the latest NICER calibration files (version 20200722).

\par The percentages of good-time-intervals (GTIs), removing the periods of high background or flaring activity, are shown in Table~\ref{tab:obs_gti}. For EPIC-pn, we adopted the standard count rate thresholds (0.4\,s$^{-1}$) to filter the event lists. We also filtered the event lists to exclude bad pixels and columns and to retain photon patterns with the highest quality energy calibration (for EPIC-pn, $\texttt{PATTERN}\le4$). The source centroid and optimal extraction region, optimizing with a high signal-to-noise ratio, were defined with the SAS task $\texttt{eregionanalyse}$ in the whole energy band from 0.2 keV to 12 keV. We ensured the choice of background regions avoided as much as possible the contamination from diffuse emission and out-of-time events. We defined background regions of sizes $40''$ to $60''$, away from the source region but on the same CCD as the target. Especially for the observations in SW mode, we verified that the results of the spectral analysis were not significantly affected by the choice of background region or window mode. Besides, we have not included ObsID 0722190201 in the spectral analysis since it is so bright that it is hard to select the background region in the same CCD.

\par For the Chandra observations, we applied the task $\texttt{chandra\_repro}$ to remove hot pixels and background flares. To minimize the contamination from diffuse emission, we optimized the source region with the help of radial profiles, e.g. by defining the extraction radius at the distance from the central target where counts rapidly drop to a near-zero level (e.g., $1.476''$ for J2323). Following \cite{Posselt2018ApJ135P}, we chose $2.46''<r<4.92''$ annuli as the background region of the J2323 and used a box to exclude the apparent filament-like structure around the center. For the data observed in CC mode, we followed the Chandra X-ray Center guidelines\footnote{\url{https://cxc.cfa.harvard.edu/ciao/caveats/acis\_cc\_mode.html}} and used box-shaped regions to select the source and background regions. 

\par For the NICER observations, the  standard NICER pipeline was applied for data reduction, with most filtering criteria chosen at their standard values. The X-ray photons in interested energy bands were selected with the task \texttt{fselect} before performing the spin frequency search. 

\par For the timing analysis, we converted the times-of-arrival (ToA) of the event files to the solar system barycenter using the $\texttt{axbary}$ tool in CIAO and $\texttt{barycen}$ in SAS and the astrometrically corrected coordinates of each target, which is especially important for observations conducted in CC mode. For NICER observations, the ToA of X-ray events were converted to the ICRS reference frame, and the ephemeris was specified to DE200.

\par To estimate the pile-up effect in XMM-Newton data, we followed the guidelines of \cite{Jethwa2015AA104J}. For the observations in FF mode, we ensured that the pile-up levels were within conservative thresholds for both the flux loss and spectral distortion (less than 1.0\% and 0.4\%, respectively). The observations conducted in SW mode are not affected by pile-up due to the higher time resolution. For Chandra, the pile-up level  estimated by the $\texttt{pileup\_map}$ tool\footnote{\url{https://cxc.harvard.edu/ciao/ahelp/pileup\_map.html}} can be considered to be overall negligible at the flux level of our targets.

\section{Analysis and Results}
\label{sect:analysis}

\subsection{Revised upper limits on the neutron star spin period}
\par We applied the $Z_{\mathrm{m}}^2$ (Rayleigh) test (\citealt{Buccheri1983AA}) directly on the times-of-arrival of the events to search for periodic signals. We can search for periodicity below Nyquist frequency, e.g., $f_{\text{max}}=1/(2t_{\text{res}})=87.71\,\text{Hz}$, where $t_{\text{res}}=5.7\,$ms is the time resolution in SW mode. But for the NICER observations, the $f_{\text{max}}$ was set to $50\,$Hz to save computing time. We set the minimum frequency to 0.01 Hz, corresponding to 100 s, to not miss the possible periodic signal since the measured pulse periods of CCOs are within 0.1 s to 0.5 s. We adopted a frequency step $f_{\text{step}}=1-4\,\mu$Hz (oversampling factor of 3 to 8) to warrant that a peak periodic signal is not missed, and the number of independent trials were listed on the Table~\ref{tab:Timing_results}. The energy range is chosen as a compromise of the total counts and the source-to-background ratio, while the determination of the radius was based on the radial profile for each target.

\par In Table~\ref{tab:Timing_results}, we summarize the results of the timing analysis for each target. For J0852, taking into account the number of independent trials (see, e.g., \citealt{Pavlov1999ApJ45P}), we found no significant pulsations with pulsed fraction above 6\% (3$\sigma$) in the $\mathrm{12\,ms-100\,s}$ range, which is consistent with the limit reported by \cite{deluca2008AIPC..983..311D}, that is, 7\% at 99\% confidence level.

\par The analysis of XMM-newton observation for J1601 does not show pulsation with a pulsed fraction larger than 18\% (3$\sigma$ threshold) in a period range of 147 ms to 100 s, which is consistent with that derived by \cite{doroshenko2018AA76D} ($<21\%$, at the same period range). 

\par Pulsation searches for J1713 yield a pulsed fraction limit of 3\% ($3\sigma$) down to a pulse period of 12\,ms, which is consistent with the limit reported by \cite{deluca2008AIPC..983..311D}, that is, less than $7\%$ ($99\%$ c.l.) in a period range of $12\,$ms to $6\,$s. 

\par The observation of J1720, performed in continuous-clocking mode, gives us the chance to search for a period above 6 ms. However, our analysis does not show any significant pulsed signal above $6\,$ms, and the pulsed fraction is no larger than 16\% (at $3\sigma$ confidence level), coincides with \cite{lovchinsky2011ApJ70L}, that is $\mathrm{PF}<16\%$ (at 99\% confidence level) in the range above 6.4 s.

\par Our analysis for J1732 does not reveal any pulsations with pulsed fraction larger than 5\% (3$\sigma$ c.l.) in the 12\,ms -- 100\,s range, which is consistent with the upper limit $\sim$8\% down to 0.2 ms at 99\% c.l. reported by \cite{klochkov2013AA}.

\par For J1818, when we reanalyzed the dataset as \cite{klochkov2016AA12K}, we did not find any pulsations ($\mathrm{PF}<50\%$, at $3\sigma$ c.l.) in the period range of $6.2\,$s to $100\,$s, consistent with the derived 99\% c.l. upper limit of 56\% reported by \cite{klochkov2016AA12K}. The $3\sigma$ upper limit using the NICER dataset (including $\sim70\%$ background) is down to 18\% in the range of $\mathrm{20\,ms-100\,s}$. 

\par Our analysis of Chandra observation for J2323 in the Cas A has not found any significant periodic signal above 0.68\,s. The $3\sigma$ confidence level value of the upper limit is 12\%, consistent with that reported by \cite{Halpern2010ApJ436H} ($\mathrm{PF}<12\%$ above $10\,$ms, 99\% c.l.).

\begin{table}[!t]
    \centering\small
    \caption{Upper limits on pulsations from timing analysis---$Z_1^2$ test}
    \label{tab:Timing_results}
    \begin{tabular}{ccccccccc}
    \hline
      Targets & ObsID & Energy Band & Photons & $f_{\text{min}}$ & $f_{\text{max}}$ & $f_{\text{step}}$ & $N_{\text{trial}}$ & PF$_{\mathrm{3\sigma}}$ \\ 
         & & keV & & Hz & Hz & $\mu\text{Hz}$ & & \% \\
    \hline
    J0852 & 0652510101 & $0.5-6.0$  & 24310 & 0.01 & 87.71 & 4 & $2.2\times10^{7}$  & 6 \\
    J1601 & 0742050101 & $1.1-6.0$  & 2457 & 0.01 & 6.81 & 1 & $6.8\times10^{6}$  & 18 \\
    J1713 & 722190101 & $0.3-6.0$  & 100590 & 0.01 & 87.71 & 1 & $8.77\times10^{7}$  & 3 \\
      & 3201030101 & $0.6-3.5$ & 38750 & 0.01 & 50 & 2 & $2.5\times10^7$ & 5 \\
    J1720 & 14806 & $1.0-6.0$  & 3656 & 0.01 & 175.4 & 2 & $8.77\times10^{7}$  & 16 \\
    J1732 & 0722190201 & $1.0-6.0$  & 44715 & 0.01 & 87.71 & 1 & $8.77\times10^{7}$  & 5 \\ 
      & 1030230107 & $1.0-5.0$ & 26830 & 0.01 & 50 & 2 & $2.5\times10^7$ & 6 \\
    J1818 & 16766 & $0.5-9.0$  & 308 & 0.01 & 0.156 & 2 & 72100  & 50 \\
          & 0034130102 & $0.5-8.0$  & 7990 & 0.01 & 50   & 2 & $2.5\times10^7$ & 18 \\ 
    J2323 & 16946 & $0.5-8.0$  & 5971 & 0.01 & 1.466 & 2 & $7.28\times10^{5}$  & 12 \\
    \hline
    \end{tabular}
    \note{$\mathrm{PF}_{3\sigma}=(2Z_{1,3\sigma}/\mathrm{N_{photons}})^{1/2}$}
\end{table}

\subsection{Spectral Analysis}
\label{sect:spec}

\par We extracted spectra from the source and background regions described in section~\ref{sect:Obs}, together with the respective response matrices and ancillary files created by SAS task \texttt{rmfgen} and \texttt{arfgen}, while these processes were included in the pipeline \texttt{specextract} for Chandra observations. For Chandra observations operated in CC mode, the \texttt{specextract} setting followed the guidelines provided by the CXC\footnote{\url{https://cxc.cfa.harvard.edu/ciao/caveats/acis\_cc\_mode.html}}. For the XMM-Newton observations, the energy channels of each spectrum were regrouped with minimum counts of at least 25 per spectral bin. We ensured the binning does not overample the instrumental energy resolution by more than a factor of 3. The counts in each group were at least 25 for the Chandra observations, except for the observation in CC mode, in which we adopted at least 30 counts/bin. To fit the spectra, we used Xspec version 12.10.1f (\citealt{Arnaud1996ASPC17A}). In the analysis, we adopted the $\texttt{tbabs}$ model and \texttt{wilm} abundance table in XSPEC (\citealt{Wilms2000ApJ914W}) to model the interstellar absorption in the line-of-sight. We estimate that the choice of abundance table and cross-section model impacts the results of the spectral fitting by 20\% -- 30\% in column density.

\par We modeled the thermal spectra of CCOs as single and multi-temperature blackbody components, absorbed by the interstellar material. Since a single blackbody model results in an unrealistic emission radius at a given distance to the source (e.g., \citealt{Pavlov2000ApJ53P,Slane2001ApJ814S,lazendic2003ApJ}), in this work we adopted a two-component blackbody model for CCOs with high enough photon statistics (except for J1818). In addition, we considered neutron star atmosphere models, $\texttt{hatm}$\footnote{\url{https://heasarc.gsfc.nasa.gov/xanadu/xspec/models/hatm.html}} (\citealt{klochkov2015AA53K,suleimanov2017AA43S}) and $\texttt{carbatm}$\footnote{\url{https://heasarc.gsfc.nasa.gov/xanadu/xspec/models/carbatm.html}} (\citealt{suleimanov2014ApJS13S}). The results for each model are in Tables~\ref{tab:bbodyrad}, ~\ref{tab:carb}, and ~\ref{tab:hatm}. For each fit in Table~\ref{tab:bbodyrad}, we list the reduced chi-squared ($\chi^2_\nu$), null-hypothesis probability (NHP in \%), the column density $n_{\text{H}}$ in units of $10^{22}\ \mathrm{cm^{-2}}$, the temperature of the cold and hot blackbody component $kT^{\infty}_{c}$, $kT^{\infty}_{h}$ in eV, and the corresponding radiation radii $R^{\infty}_{c}$, $R^{\infty}_{h}$. 
When fitting the data with a two-component hydrogen atmosphere model, we linked the normalization of both components through an additional factor $\delta$, where $0\le\delta\le1$ was defined by the fraction of the surface area occupied by the hot spots (\citealt{suleimanov2017AA43S}). The normalization in $\texttt{carbatm}$ model is defined as $N=A/D_{\mathrm{10kpc}}^2$, where ``A" characterizes the fraction of the surface area emitting the radiation and $D_{\mathrm{10kpc}}$ is the distance to the source in units of 10 kpc. We kept the mass and radius fixed at default values in the fitting procedure, while the ``A" parameter was free to vary. For most fits, the column density varies between $0\,cm^{-2}$ and $5\times10^{22}\,cm^{-2}$. While for J1601, J1720, and J1818, the column density varies between $0\,cm^{-2}$ and $10\times10^{22}\,cm^{-2}$.  We discuss in the remainder of this section the results for each source individually.

\begin{table}[!t]
    \centering\small
    \caption{The best-fit results for the two-component blackbody model}
    \label{tab:bbodyrad}
    \begin{tabular}{cccccccccc}
    \hline
     Target & EB & $\chi_{\nu}^2$(d.o.f) & NHP  & $n_{\text{H}}$ & $kT^{\infty}_{c}$ & $R^{\infty}_{c}$ & $kT^{\infty}_{h}$ & $R^{\infty}_{h}$ & Flux$^{a}$ \\ 
     & $\mathrm{keV}$ & & \%  & $10^{22}\ \mathrm{cm^{-2}}$ & $\mathrm{eV}$ & $\mathrm{km}$ & $\mathrm{eV}$ & $\mathrm{km}$ &  \\
    \hline
    J0852 & 0.8-5.0 & 0.87(68) & 77 & $0.55_{-0.05}^{+0.06}$ & $317_{-44}^{+33}$ & $0.39_{-0.04}^{+0.08}\,d_1$ & $482_{-43}^{+89}$ & $0.12_{-0.06}^{+0.07}\,d_1$ & 1.34(1) \\
    J1601 & 1.0-6.0 & 1.01(47) & 45 & $4.85_{-0.65}^{+0.84}$ & $303_{-67}^{+84}$ & $1.41_{-0.67}^{+2.23}\,d_{4.9}$ & $531_{-55}^{+150}$ & $0.28_{-0.18}^{+0.10}\,d_{4.9}$ & 0.124(3) \\
    J1713 & 0.4-5.0 & 0.95(92) & 61 & $0.55_{-0.01}^{+0.01}$ & $343_{-11}^{+10}$ & $0.67_{-0.02}^{+0.02}\,d_{1.3}$ &  $552_{-24}^{+30}$ & $0.17_{-0.03}^{+0.03}\,d_{1.3}$ & 3.181(12) \\
    J1720 & 0.5-8.0 & 0.83(115) & 90 & $4.97_{-0.65}^{+1.17}$ & $233_{-88}^{+133}$ & $2.17_{-1.32}^{+10.36}\,d_{4.5}$ & $511_{-11}^{+26}$ & $0.72_{-0.15}^{+0.03}\,d_{4.5}$ &  0.49(1) \\
    J1732$^{b}$ & 0.5-6.0 & 1.25(192) & 1 & $2.19_{-0.05}^{+0.06}$ & $426_{-23}^{+18}$ & $1.30_{-0.08}^{+0.10}\,d_{3.2}$ & $688_{-64}^{+99}$ & $0.24_{-0.10}^{+0.13}\,d_{3.2}$ & 2.664(15) \\
    J2323 & 0.8-6.0 & 0.95(132) & 64 & $1.92_{-0.23}^{+0.17}$ & $286_{-48}^{+63}$ & $1.49_{-0.51}^{+1.15}\,d_{3.4}$ & $479_{-28}^{+59}$ & $0.45_{-0.19}^{+0.13}\,d_{3.4}$ & 0.63(1) \\

    \hline
    \end{tabular}
    \note{The model fitted to the data in XSPEC is $\texttt{TBabs*(bbodyrad+bbodyrad)}$. Errors are 1$\sigma$ confidence levels. $d$ with subscript n.m is the distance in n.m kpc (e.g., $d_{1.3}$ is the distance in 1.3 kpc). $^{a}$ The observed flux is in units of $10^{-12}\ \mathrm{erg\ s^{-1}cm^{-2}}$ in energy band 0.5 keV -- 12 keV, and the errors are shown in parenthesis. $^{b}$ The simultaneous fit of EPIC-pn (ObsID 0722090101 and 0694030101).}
\end{table}

\paragraph{\bf{CXOU J085201.4-461753}} We fit the spectra of the CCO J0852 in the SNR G266.2-1.2 (aka Vela Jr.) ($2.4-5.1\,$kyr, \citealt{Pavlov2001ApJ131P,kargaltsev2002ApJ,Allen2015ApJ82A}) in the energy range of 0.8 keV to 5.0 keV to minimize the influence of the remnant Vela in the line of sight. The two-temperature blackbody model gives $kT^{\infty}_{c}=317_{-44}^{+33}$ eV with $R^{\infty}_{c}=0.39_{-0.04}^{+0.08}$ km and $kT^{\infty}_{h}=482_{-43}^{+89}$ eV with $R^{\infty}_{h}=0.12_{-0.06}^{+0.07}$ km, where $d_{1}=1\,$kpc provided by \cite{Allen2015ApJ82A}. The hot-spot in the two-component model also contributes most flux to the spectrum so that the colder surface component is insignificant. Assuming a third blackbody component, we obtain $kT^{\infty}_{c}<89$ eV, which agrees with that derived by Danilenko et al. ($90\pm10$ eV; \citealt{Potekhin2020MNRAS5052P}). 

For the atmosphere models with canonical mass and radius, the best-fit distance for the $\texttt{carbatm}$ model is about 2.5 kpc, which is higher than that estimated for G266.2-1.2 \cite{Allen2015ApJ82A}. Similar results are obtained for the two-component $\texttt{hatm}$ model (Table~\ref{tab:hatm}). The best-fit column density in the direction to the source $(0.7-0.8)\times10^{22}\,\mathrm{cm^{-2}}$ agrees to that derived for G266.2-1.2 ($<1.1\times10^{22}\,\mathrm{cm^{-2}}$; \citealt{Slane2001ApJ814S,Acero2013AA7A,Allen2015ApJ82A}). 

\paragraph{\bf{CXOU J160103.1-513353}} The CCO J1601 in the SNR G330.2+1.0 ($1\,$kyr, \citealt{Borkowski2018ApJ21B}) is located at least $4.9\pm0.3$ kpc away from Earth (\citealt{McClure2001ApJ394M}). Its age, estimated from the decelerated expansion of the associated SNR, is younger than 1 kyr (\cite{Borkowski2018ApJ21B}). Using a two-component blackbody model, the temperature of the cool and hot components are $303_{-67}^{+84}$ eV and $531_{-55}^{+150}$ eV, corresponding to radiation radii of $1.41_{-0.67}^{+2.23}\,d_{4.9}$km and $0.28_{-0.18}^{+0.10}\,d_{4.9}$ km, respectively. The upper limit on the temperature of a cold surface is  $kT_{c}^{\infty}<138\,$eV, which is consistent with that presented by \cite{doroshenko2018AA76D}, using the carbon atmosphere model. 

Regarding atmosphere models, both the carbon and two-component hydrogen atmospheres describe the data well. However, as discussed by \cite{doroshenko2018AA76D}, the both models require a higher absorption column density than that derived for G330.2+1.0 ($n_{\text{H}}\le3.15\times10^{22}\,\mathrm{cm^{-2}}$; \citealt{Williams2018ApJ118W}). The distance assuming radiation from the whole neutron star surface is $6.38_{-1.39}^{+1.67}$ kpc for the \texttt{carbatm} model, which coincides with the lower limit of $4.9\pm0.3$ kpc derived by \cite{McClure2001ApJ394M}. Even though none of the atmosphere models is preferred, the hot-spot contributes more than 69\% flux in the energy band of 0.5 keV to 10 keV, which may mean that the cool component is not significant for the fit. This is in line with the results of \cite{doroshenko2018AA76D} and \cite{park2009ApJ431P}.

\begin{table}[!t]
    \centering\small
    \caption{Results of the best-fit three-component blackbody model}
    \label{tab:fit3bb}
    \begin{tabular}{ccccccccccc}
    \hline
    Targets & $\chi_{\nu}^2$ & NHP  & $n_H$ & $kT^{\infty}_{\mathrm{c}}$ & $R^{\infty}_{\mathrm{c}}$ & $kT_{\mathrm{h1}}$ & $R^{\infty}_{\mathrm{h1}}$ & $kT_{\mathrm{h2}}$ & $R^{\infty}_{\mathrm{h2}}$ & Flux \\
    &   & \% & $10^{22}\mathrm{cm^{-2}}$ & $\mathrm{eV}$ & $\mathrm{km}$ & $\mathrm{eV}$ & $\mathrm{km}$ & $\mathrm{eV}$ & $\mathrm{km}$ &  \\ \hline
    J0852 & 0.83 & 83 & $0.92_{-0.30}^{+0.28}$ & $89_{-16}^{+13}$ & $13\,d_1$ & $243_{-22}^{+58}$ & $0.77_{-0.33}^{+0.26}\,d_1$ & $438_{-17}^{+42}$ & $0.20_{-0.06}^{+0.03}\,d_1$ & 1.341(11) \\
    J1601 & 1.0 & 46 & $5.58_{-1.25}^{+0.89}$ & $138^{+26}$ & $15.1\,d_{4.9}$ & $289_{-33}^{+56}$ & $1.86_{-0.97}^{+1.89}\,d_{4.9}$ & $526_{-51}^{+122}$ & $0.29_{-0.16}^{+0.10}\,d_{4.9}$ & 0.124(3) \\
    J1713 & 0.89 & 75 & $0.89\pm0.03$ & $95\pm3$ & $15.1\,d_{1.3}$ & $266\pm7$ & $1.21_{-0.05}^{+0.06}\,d_{1.3}$ & $486_{-9}^{+10}$ & $0.30_{-0.02}^{+0.02}\,d_{1.3}$ & 3.173(12) \\
    J1732 & 1.19 & 4.1 & $2.34_{-0.09}^{+0.11}$ & $115_{-10}^{+9}$ & $15.1\,d_{3.2}$ & $402_{-25}^{+22}$ & $1.46_{-0.12}^{+0.15}\,d_{3.2}$ & $643_{-44}^{+66}$ & $0.33_{-0.11}^{+0.12}\,d_{3.2}$ & 2.660(15) \\
    J2323 & 0.96 & 63 & $2.0_{-0.3}^{+0.4}$ & $101^{+23}$ & $13.4\,d_{3.4}$ & $281_{-44}^{+56}$ & $1.61_{-0.29}^{+0.90}\,d_{3.4}$ & $479_{-27}^{+52}$ & $0.46_{-0.18}^{+0.12}\,d_{3.4}$ & 0.63(1) \\
    
    \hline
    \end{tabular}
    \note{The model fitted to the data in XSPEC is $\texttt{TBabs(bbodyrad+bbodyrad+bbodyrad)}$. Errors are 1$\sigma$ confidence levels. The observed model flux is in units of $10^{-12}\mathrm{erg\ s^{-1}\ cm^{-2}}$ in energy band 0.5 -- 12 keV.}
\end{table}

\paragraph{\bf{1WGA J1713.4-3949}} The CCO J1713 in the SNR G347.3-0.5 ($\sim1.6\,$kyr, \citealt{Slane1999ApJ357S,lazendic2003ApJ,wangZR1997AA}) is located $1.3\pm0.4$ kpc away from the observer (\citealt{cassam2004aAA}). The double-blackbody (2BB) model fit to the spectrum of the source gives $kT^{\infty}_{c}=343_{-11}^{+10}$ eV, $R^{\infty}_{c}=0.67_{-0.02}^{+0.02}\,d_{1.3}$ km for the cold component and $kT^{\infty}_{h}=552_{-24}^{+30}$ eV, $R^{\infty}_{h}=0.17_{-0.03}^{+0.03}\,d_{1.3}$ km for the hot component, where $d_{1.3}$ is the distance in units of 1.3 kpc (\citealt{cassam2004aAA}). If we assume the emission from the colder neutron star surface is absorbed by the high Galactic absorption, we obtain an upper limit $kT_{c}^{\infty}\le95$ eV for $R_{c}^{\infty}=15.1\,d_{1.3}\,$km (see Table~\ref{tab:fit3bb}), which is more acceptable for a young (e.g., $10^3-10^4\,$years) cooling neutron star ($<90-110\,$eV, assuming standard cooling for a canonical neutron star, \citealt{yakovlev2004ARAA169Y}). By comparison, \cite{Potekhin2020MNRAS5052P} gave a value of $138\pm1$ eV using atmosphere model ($\texttt{TBabs*NSX}$).

\par The fit of atmosphere models with fixed neutron star mass and radius, 1.5 $M_{\odot}$ and 12 km, respectively, have a $\chi_{\nu}^{2}=0.98$ for \texttt{carbatm} and $\chi_{\nu}^2=1.3$ for a two-component \texttt{hatm} model. The fraction of the emitting area is about 0.2 at a fixed distance of 1.3 kpc. If we assume the radiation arises from the whole surface, the distance to the source is $2.88_{-0.07}^{+0.06}$ kpc (at 99\% confidence level), which is inconsistent with that derived for G347.3-0.5. Alternatively, the fit with free mass and radius for a source at a distance of 1.3 kpc gives best-fit parameters 0.66 $M_{\odot}$ and 7.1 km, overall inconsistent with nucleonic matter (\citealt{ozel2016ARAA..54..401O}).

\begin{table}[!t]
    \centering
    \caption{Results of the best-fit carbon atmosphere model}
    \label{tab:carb}
    \begin{tabular}{ccccccccc}
    \hline
      CCO & $\chi_{\nu}^2$ & $\mathrm{NHP}$ & $n_{\mathrm{H}}$ & $\mathrm{T}$ & $A$ & Flux \\
      & & \%  & $10^{22}\mathrm{cm^{-2}}$ & $\mathrm{MK}$ & &  \\
    \hline
     J0852 & 0.86 & 79 & $0.70_{-0.02}^{+0.02}$ & $1.68_{-0.03}^{+0.03}$ & 0.13 & 1.34(1) \\
     J1601 & 0.98 & 51 & $4.71_{-0.26}^{+0.25}$ & $1.84_{-0.12}^{+0.13}$ & 0.59 & 0.124(3) \\
     J1713 & 0.98 & 55 & $0.71_{-0.01}^{+0.01}$ & $1.97_{-0.02}^{+0.01}$ & 0.2  & 3.185(12) \\
     J1720 & 0.89 & 80 & $5.74_{-0.23}^{+0.24}$ & $2.37_{-0.10}^{+0.11}$ & 0.9  & 0.50(1) \\
     J1732 & 1.32 & 0.18 & $2.57_{-0.03}^{+0.03}$ & $2.32_{-0.03}^{+0.03}$ & 0.81 & 2.656(15)\\
     J2323 & 0.95 & 66 & $2.06_{-0.08}^{+0.09}$ & $1.97_{-0.07}^{+0.07}$ & 0.92 & 0.63(1) \\
    \hline
    \end{tabular}
    \note{Errors are 1$\sigma$ confidence levels. ``A" characterizes the fraction of the surface area emitting the radiation. The observed flux is in unit of $10^{-12}\ \mathrm{erg\ s^{-1}cm^{-2}}$ in energy band 0.5 keV -- 12 keV.}
\end{table}

\paragraph{\bf{XMMU J172054.5-372652}} The CCO J1720 associated with the SNR G350.1-0.3 ($\sim1\,$kyr), which is 4.5 kpc away from observer (\citealt{gaensler2008ApJ37G,lovchinsky2011ApJ70L}). Both single and double blackbody/hydrogen atmosphere models provide statistically acceptable fit results. The $\chi_{\nu}^2$ values are within 0.8 to 0.9 (d.o.f 115). The absorption column density of $(4.5-5.0)\times10^{22}\,\mathrm{cm^{-2}}$ is a factor of $0.2-0.3$ higher than that derived by \cite{gaensler2008ApJ37G} and \cite{lovchinsky2011ApJ70L}. If we only consider a single temperature model -- for instance, the blackbody model -- the emitting area is about 0.7 $d_{4.5}\,$ km at a distance of 4.5 kpc (\citealt{gaensler2008ApJ37G}). Emission from the entire surface is inconsistent with a source located in the Milky Way. The temperature of 0.5 keV is considerably higher than that expected ($<90\,$eV, assuming standard cooling curve by \cite{yakovlev2004ARAA169Y}) for a young NS of $\sim0.9$ kyr. In comparison, the carbon atmosphere model provides good fits with a colder temperature of $2.37_{-0.10}^{+0.11}$ MK at a distance of $4.74_{-0.71}^{+0.77}$ kpc, consistent with that measured for the SNR (\citealt{gaensler2008ApJ37G}). However, the model requires a higher absorption column density of $5.74_{-0.23}^{+0.24}\times10^{22}\,\mathrm{cm^{-2}}$ than the simple blackbody and power-law model (\citealt{lovchinsky2011ApJ70L}) to describe the emitted spectrum. This may indicate another absorption component in the line of sight.

\paragraph{\bf{XMMU J173203.3-344518}} The CCO J1732 in the SNR G353.6-0.7 ($2-6\,$kyr, \citealt{halpern2010ApJ941H,Acero2015AA74A}) is located at a distance of $d\approx3.2\,\mathrm{kpc}$ (\citealt{Maxted2018MNRAS662M}). 
The fit of a two-component blackbody model has $\chi_{\nu}^{2}=1.25$ (192 d.o.f.). And the best-fit $kT^{\infty}_{c}=426_{-23}^{+18}\,$eV, $kT^{\infty}_{h}=688_{-64}^{+99}\,$eV, corresponding with $R^{\infty}_{c}=1.3\pm0.1\,d_{3.2}\,$km and $R^{\infty}_{h}=0.24^{+0.13}_{-0.1}\,d_{3.2}\,$km. The inclusion of a third blackbody component consistent with emission from the neutron star surface has $kT<115$\,eV (Table~\ref{tab:fit3bb}). For neutron star atmosphere models with canonical mass and radius ($1.5\,M_{\odot}$ and $12\,$km) at a distance of 3.2 kpc, we obtain $\chi_{\nu}=1.32$. Using the $\texttt{carbatm}$ model, the emitting fraction is about 0.8, while the column density of $n_{H}=2.6\times10^{22}\,\mathrm{cm^{-2}}$ is inconsistent with that derived by \cite{klochkov2013AA,klochkov2015AA53K} ($\sim2.0\times10^{22}\,\mathrm{cm^{-2}}$), which may be caused by the selection of the abundance table.

\begin{table}[htb]
    \centering
    \caption{Results of the best-fit two-temperature hydrogen atmosphere model}
    \label{tab:hatm}
    \begin{tabular}{ccccccccc}
    \hline
      CCO & $\chi_{\nu}^2$ & $\mathrm{NHP}$ & $n_{\mathrm{H}}$ & $\mathrm{T_1}$ & $\mathrm{T_2}$ & $\delta$ & Flux \\
      & & \%  & $10^{22}\ \mathrm{cm^{-2}}$ & $\mathrm{MK}$ & $\mathrm{MK}$ & &  \\
    \hline
     J0852 & 0.88 & 76 & $0.796_{-0.110}^{+0.067}$ & $0.82_{-0.11}^{+0.06}$ & $2.97_{-0.06}^{+0.07}$ & 0.0063 & 1.332(11) \\
     J1601 & 0.99 & 49 & $5.30_{-0.40}^{+0.33}$ & $1.62_{-0.17}^{+0.10}$ & $4.11_{-0.46}^{+0.56}$ & 0.0061 & 0.124(3) \\
     J1713 & 1.31 & 2.2 & $0.71_{-0.02}^{+0.01}$ & $0.96_{-0.04}^{+0.03}$ & $3.33_{-0.03}^{+0.03}$ & 0.0143 & 3.163(12) \\
     J1720 & 0.82 & 91 & $5.03_{-0.25}^{+0.50}$ & $1.15^{+0.6}$ & $4.34_{-0.13}^{+0.13}$ & 0.0402 & 0.497(8) \\
     J1732 & 1.39 & $10^{-2}$ & $2.62_{-0.07}^{+0.06}$ & $1.79_{-0.09}^{+0.07}$ & $4.25_{-0.10}^{+0.11}$ & 0.0369 & 2.647(15)\\
     J2323 & 0.95 & 65 & $2.12_{-0.18}^{+0.14}$ & $1.62_{-0.23}^{+0.13}$ & $3.85_{-0.26}^{+0.30}$ & 0.0239 & 0.63(1) \\

    \hline
    \end{tabular}
    \note{Errors are 1$\sigma$ confidence levels. Where $0\le\delta\le1$ is defined by the fraction of the surface area occupied by the hot-spots. The observed flux is in unit of $10^{-12}\ \mathrm{erg\ s^{-1}cm^{-2}}$ in energy band 0.5 keV -- 12 keV.}
\end{table}

\paragraph{\bf{CXOU J232327.9+584842}} The CCO J2323 in Cas A, one of the youngest known neutron stars (\citealt{Ashworth1980JHA1A}, \citealt{fesen2006ApJ...636..848F}), is located at a distance of 3.4 kpc (\citealt{Reed1995ApJ706R}). The best-fit 2BB model has temperatures and radii of $286_{-48}^{+63}\,$eV ($R^{\infty}_{c}=1.49_{-0.51}^{+1.15}\,d_{3.4}\,$km) and $479_{-28}^{+59}$ eV ($R^{\infty}_{h}=0.45_{-0.19}^{+0.13}\,d_{3.4}\,$km). If arising from hot spots on the surface, the thermal components contribute the most flux to the source luminosity. Adding a third component to the blackbody model, we find an upper limit of the cold component of $kT_{c}^{\infty}<124$ eV, which is consistent with the value of $123-185\,$eV from \citet{Wijngaarden2019MNRAS974W,heinke2010ApJ167H}.

The spectrum of J2323 can also be well described by the carbon atmosphere model (\cite{Ho2009Natur71H}). In our analysis we assumed the mass and radius of 1.647 $\mathrm{M_{\odot}}$ and 10.33 km, respectively, from \cite{Posselt2013ApJ186P}. The column density agrees with that derived for the 2BB model (Table~\ref{tab:carb}; see also \citealt{Posselt2018ApJ135P}) but is a bit higher than that reported by \cite{Ho2009Natur71H}. This is likely due to the different adopted abundance tables.

\paragraph{\bf{CXOU J181852.0-150213}} The CCO J1818, associated with the SNR G15.9+0.2 ($3-6\,$kyr), is located at a distance of $8.5-16\,$kpc (\citealt{reynolds2006ApJ45R,Sasaki2018MNRAS3033S}). 
The best-fit results are presented in Table~\ref{tab:1818fit}. All of the tested models provide acceptable fits to the data given the low photon statistics. While the hydrogen column density from both the blackbody and hydrogen atmosphere models is consistent with that measured for the SNR (\citealt{klochkov2016AA12K}), the distance is $1-2$ orders of magnitude larger than its expected range, $8.5-16.7$ kpc (\citealt{reynolds2006ApJ45R,Caswell1982MNRAS1143C,Sasaki2018MNRAS3033S}). By contrast, the carbon atmosphere model gives a consistent distance of $16_{-9}^{+20}$ kpc in general agreement with the supernova remnant (\citealt{reynolds2006ApJ45R,Sasaki2018MNRAS3033S}). In addition, compared with other models, the temperature ($\sim170\,$eV) given by the carbon atmosphere model is more consistent with that estimated by the standard cooling model, which should be within 110 eV.

\begin{table}[htb]
    \centering\small
    \caption{The best-fit results of J1818}
    \label{tab:1818fit}
    \begin{tabular}{c|ccc}
    \hline
    \diagbox{Param.}{Model} & $\texttt{bbodyrad}$ & $\texttt{hatm}$ & $\texttt{carbatm}$ \\ 
    \hline
    $n_{\mathrm{H}}(10^{22}\ \mathrm{cm^{-2}})$ & $3.6_{-1.0}^{+1.1}$ & $4.2_{-1.1}^{+1.2}$ & $5.1_{-1.2}^{+1.2}$ \\
    T$(\mathrm{MK})$ & $5.7\pm0.6$ & $3.9\pm0.6$ & $2.0\pm0.5$ \\
    D$(\mathrm{kpc})$ & $295_{-94}^{+132}$ & $74_{-31}^{+50}$ & $16_{-9}^{+20}$ \\
    $\chi_{\mathrm{red}}^{2}(\mathrm{d.o.f})$ & 0.61(10) & 0.59(10) & 0.58(10) \\
    Flux & \multicolumn{3}{c}{$\sim3.2(4)\times10^{-14}\ \mathrm{erg\ s^{-1}cm^{-2}}$} \\ 
    \hline
    \end{tabular}
    \note{Errors are $1\sigma$ confidence levels. The observed flux is in unit of $10^{-12}\ \mathrm{erg\ s^{-1}cm^{-2}}$ in energy band 0.5 keV -- 12 keV.}  
\end{table}

\section{Discussion}
\label{sect:discussion}

\par The temperature on the neutron star surface is expected to vary for magnetic field intensities higher than $10^{10}$ G. The heat conduction perpendicular to the magnetic field decreases while it increases along the field lines \citep{Greenstein1983ApJ...271..283G}. Furthermore, strong crustal fields ($B>10^{12}$ G) lead to a highly anisotropic surface temperature variation (\citealt{geppert2004AA267G}).

\par A single-component blackbody model cannot describe the spectrum of CCOs. Additional thermal components may closely characterize the spectrum. However, the obtained radii of emitting areas for cold components are ridiculous for a neutron star if we assume the hot component corresponds to the tiny hot-spot region. Here we attempt to assume a three-component blackbody model to investigate the viability of the hot-spot model in CCOs with no detected pulsations.

\subsection{The viewing geometry and presence of hot-spots}
\label{sect:geo}

\par The observed pulsed fraction depends on the geometrical configuration of the hot spots and the angle between the neutron star spin axis and the line of sight, which has been discussed by e.g., \cite{Elshamouty2016ApJ162E} for the X-ray binaries and \cite{suleimanov2017AA43S} for CCOs. Strong flux modulations are observed when the rotating axis misaligns with the line of sight. For sources with no detected periodicities, the limits on the pulsed fraction can be used to constrain the viewing geometry assuming the multi-temperature spectral model.

\par Here we considered the model of a slowly rotating neutron star observed at an inclination angle $i$ with two polar hot-spots located at an angle $\theta_{B}$ concerning the spin axis (see, e.g., \citealt{Schwope2005AA597S,Suleimanov2010AA,Pires2019AA73P}, for a full description). The light bending near the surface depends on the compactness of the neutron star, which is described by \cite{Beloborodov2002ApJ85B} (the ratio between the neutron star and Schwarzschild radius is $r_{g}=R_{ns}c^2(2GM_{ns})^{-1}$). 

\par The hot-spot model applied to the best-fit two-component blackbody model results in small pulsed fractions, whatever in which angle to observe, smaller than the upper limits derived in timing analysis, which may imply a nearly homogeneous temperature distribution for all the investigated targets. Therefore, we adopted the parameters of the three-component blackbody model (Table~\ref{tab:fit3bb}) and assumed two hot-spot with different temperatures and emitting areas located diametrically opposed to each other (hereafter, the `ad-hoc'' model). We also applied a variant of the model where the temperature profile is assumed to follow that of \cite{geppert2004AA267G} (hereafter, the ``crustal'' model). 

\par In the following Table~\ref{tab:geo_par3bb}, we list all parameters applied to the light-curve modeling process assuming a three-component blackbody model. $T_{c}$ represents the temperature of the whole cool surface, while $T_{h1}$ and $T_{h2}$ are the temperatures of the hot spots, corresponding with the hot-spot size $\theta_{sp1}$ and $\theta_{sp2}$, respectively. Moreover, the spectral analysis in Section~\ref{sect:spec} gave the absorbed column density. We did not include the analysis of J1818 since we did not get an acceptable result with a multi-component model for poor statistic photons. 

\begin{table}[htb]
    \centering\small
    \caption{parameters for each target assuming 3-BB model}
    \label{tab:geo_par3bb}
    \begin{tabular}{cccccccccc}
    \hline
     Target & $n_{\mathrm{H}}$ & $kT_{\mathrm{c}}$ & $kT_{\mathrm{h1}}$ & $\theta_{\mathrm{sp1}}$ & $kT_{\mathrm{h2}}$ & $\theta_{\mathrm{sp2}}$ & $M_{\odot}$ & $R^{\infty}$ &  $R/r_{g}$ \\ 
     & $10^{20}\ \mathrm{cm^{-2}}$ & $\mathrm{eV}$ & $\mathrm{eV}$ & $(^{\circ})$ & $\mathrm{eV}$ & $(^{\circ})$ & & $\mathrm{km}$  & \\
    \hline
    J0852 & 99.19 & 88.6 & $243.5$ & 3.39 & $437.8$ & 0.86 & 1.5 & 15.1 & 2.71  \\
    J1601 & 553 & 136.65 & $289.1$ & 7.0 & $526.3$ & 1.1 & 1.5 & 15.1 & 2.71 \\
    J1713 & 88.53 & 95.06 & $265.95$ & 4.6 & $486$ & 1.13 & 1.5 & 15.1 & 2.71 \\ 
    J1732 & 233.8 & $115$ & $402.2$ & 5.51 & 643 & 1.23 & 1.5 & 15.1 & 2.71 \\ 
    J2323 & 201  & $101.5$ & 280.9 & 6.89 & $479.4$ & 1.95 & 1.647 & 14.2 & 2.124 \\
    \hline
    \end{tabular}
\end{table}

\par We integrate the photon flux originating from the visible area of the neutron star surface for a particular viewing geometry (i, $\theta_{B}$), assuming blackbody emission and considering the light-bending (\citealt{Beloborodov2002ApJ85B}). Moreover, we also take the interstellar absorption into consideration to correct the actual flux from the surface. Furthermore, the absorption was then folded with the EPIC-pn response to give the computed count rates at a given energy band, while the energy band was given in Table~\ref{tab:bbodyrad}. We created a grid of light curves for (i, $\theta_{B}$) within ($0^{\circ}$, $0^{\circ}$) and ($90^{\circ}$, $90^{\circ}$) and computed the pulsed fraction for each orientation as $\mathrm{PF}=\frac{CR_{\mathrm{max}}-CR_{\mathrm{min}}}{CR_{\mathrm{max}}+CR_{\mathrm{min}}}$.

\par
\begin{figure}[!t]
\centering
\subcaptionbox{J0852}{
    \includegraphics[width=0.3\linewidth]{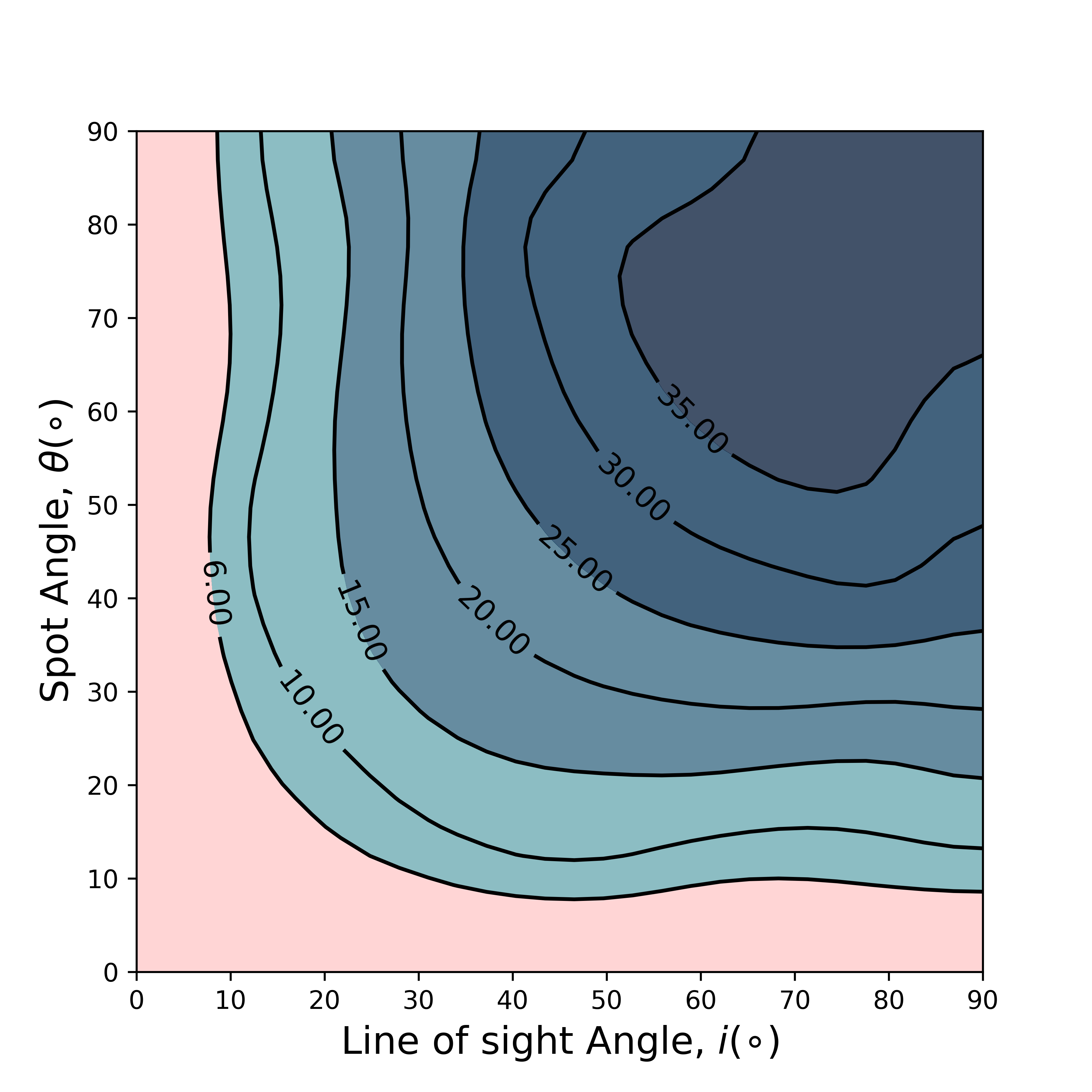}
}
\subcaptionbox{J1601}{
    \includegraphics[width=0.3\linewidth]{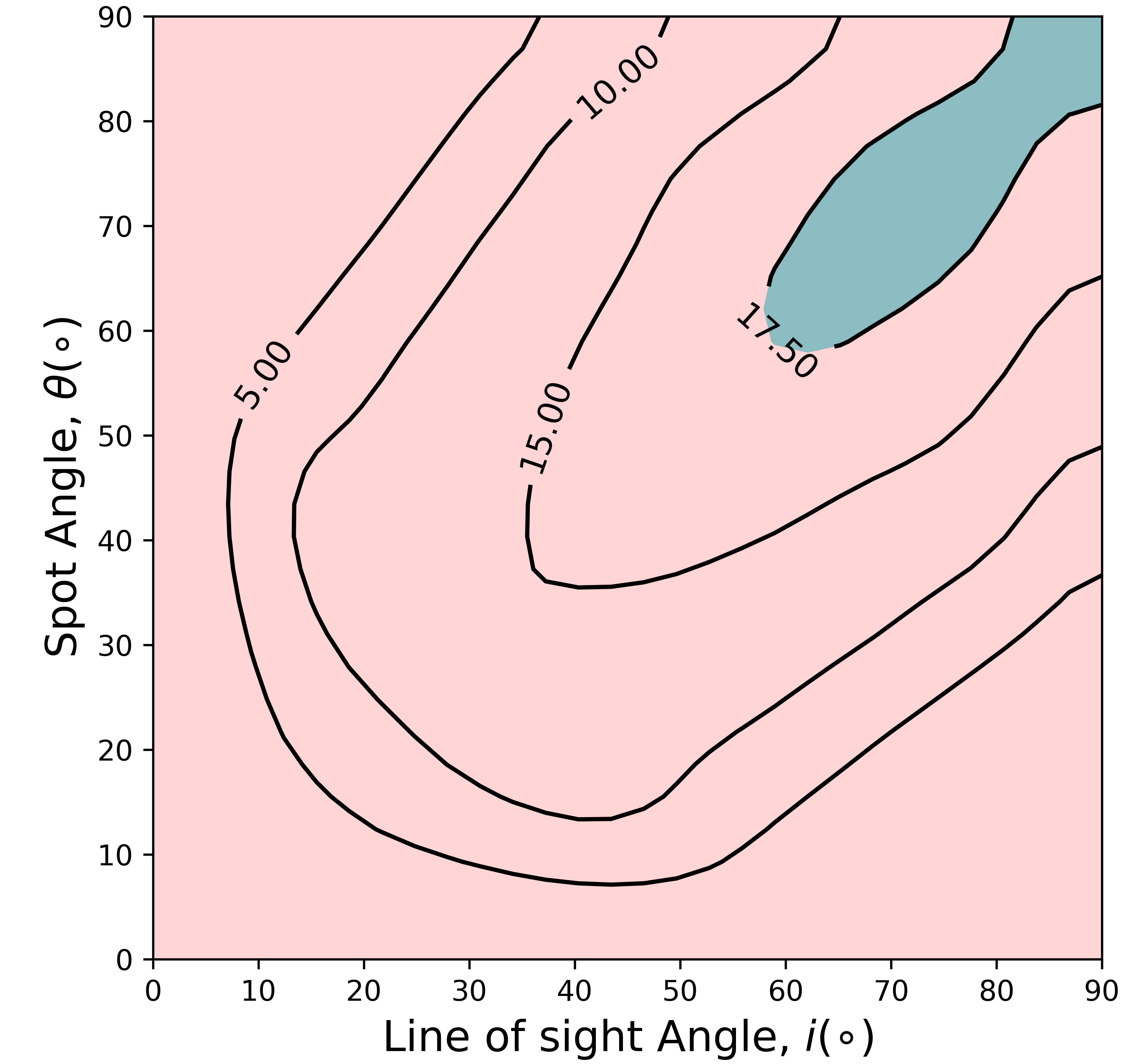}\quad
}
\subcaptionbox{J1713}{
    \includegraphics[width=0.3\linewidth]{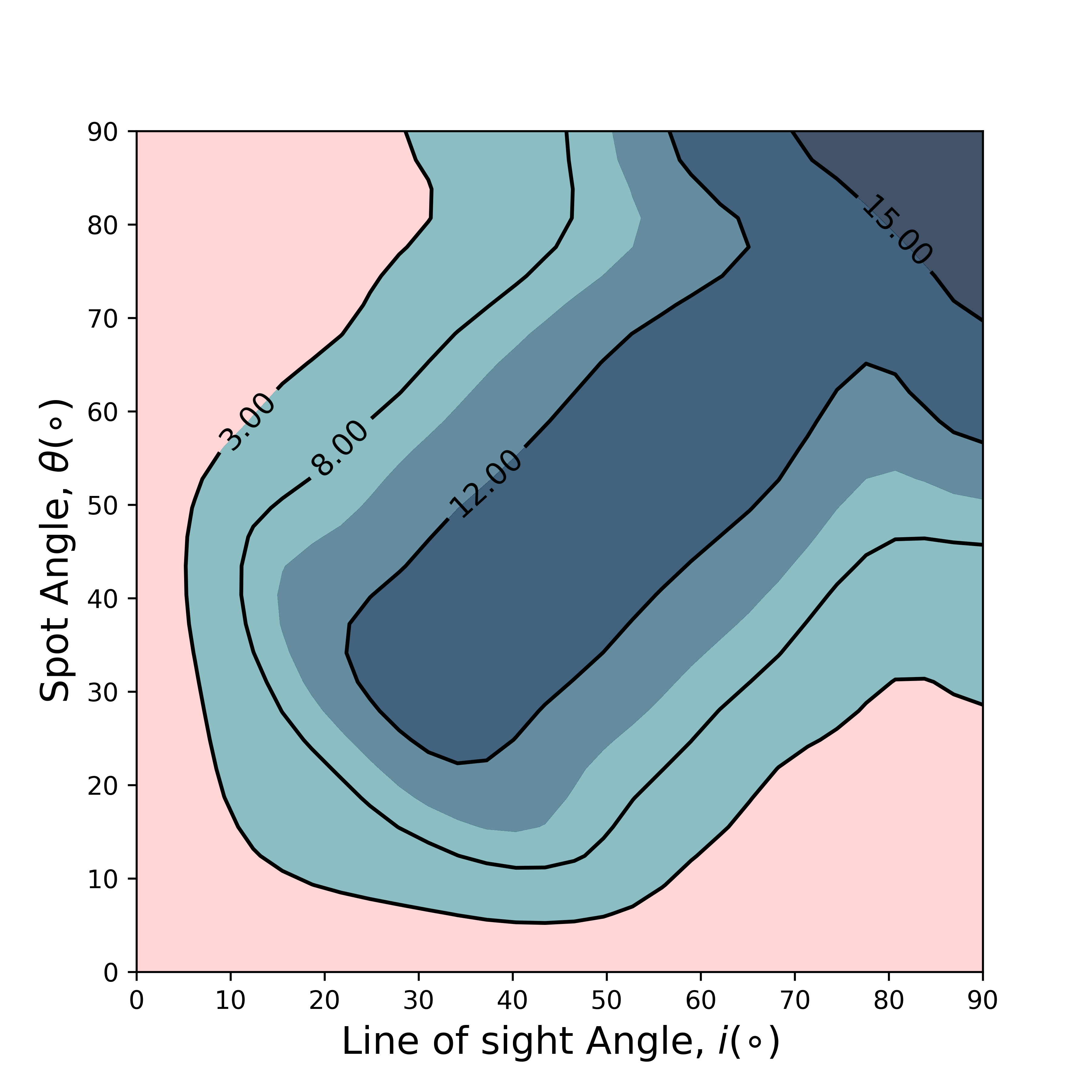}\quad
}
\smallskip
\subcaptionbox{J1732}{
    \includegraphics[width=0.3\linewidth]{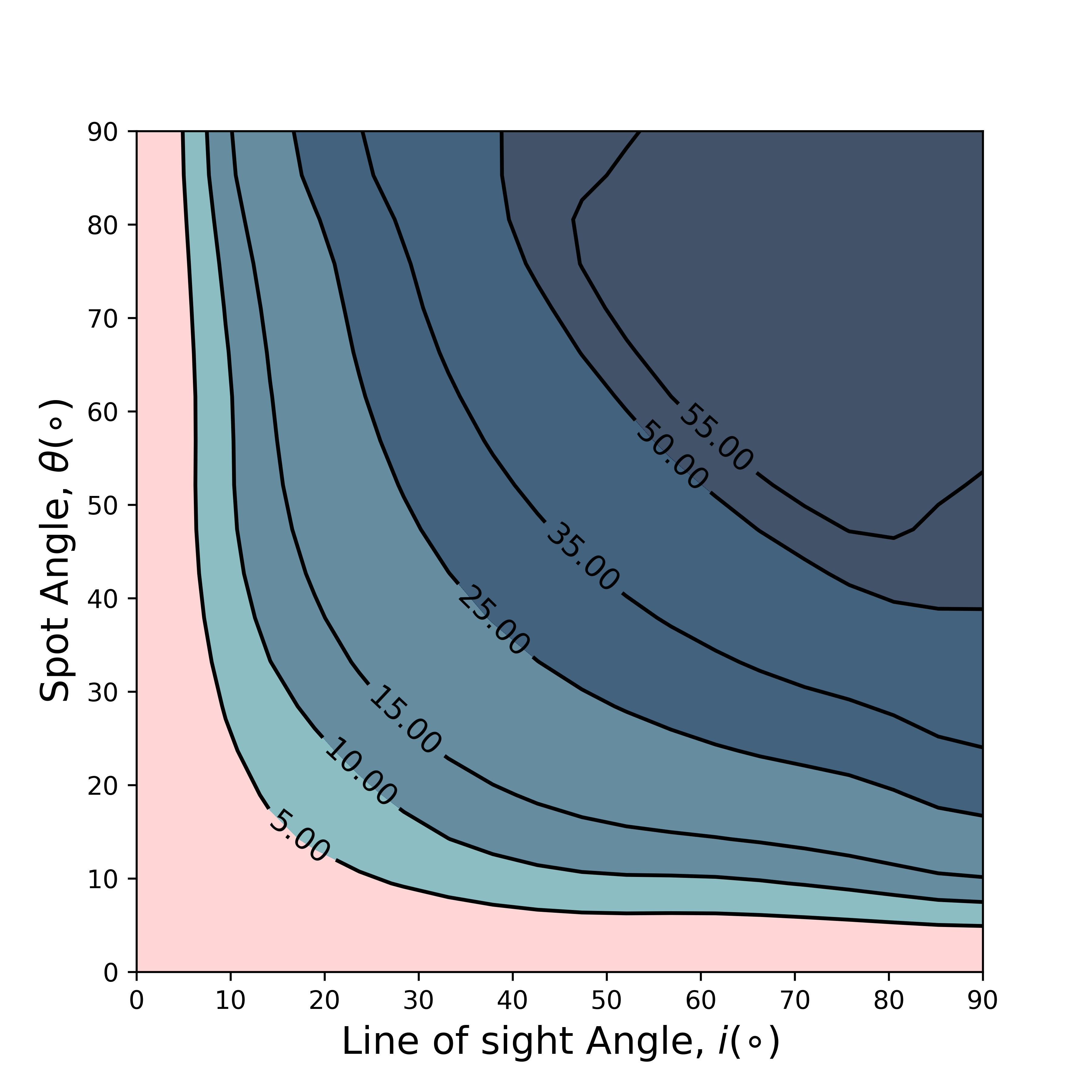}
}
\subcaptionbox{J2323}{
    \includegraphics[width=0.3\linewidth]{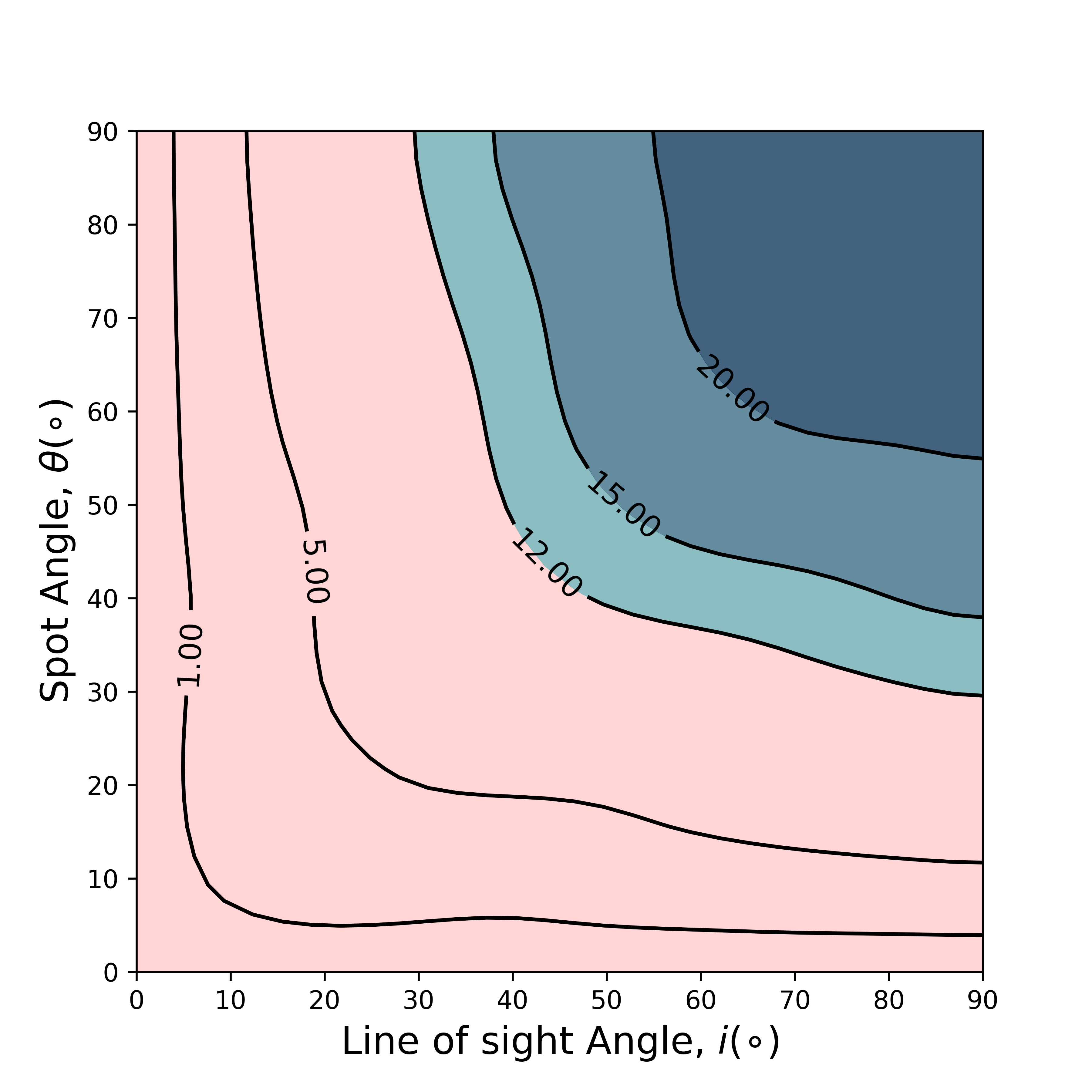}
}

\caption{Contours of constant pulsed fraction in the $(i,\theta_{\mathrm{B}})$ plane assuming the best-fit parameters of the 3-BB model. The pulsed fraction labels are given in \%. The allowed regions within the pulsed fraction limits are shown in the pink region. }
\label{fig:pfmap_3bb}
\end{figure}

\begin{figure}[!t]
\centering\small
\subcaptionbox{J0852}{
   \includegraphics[width=0.3\linewidth]{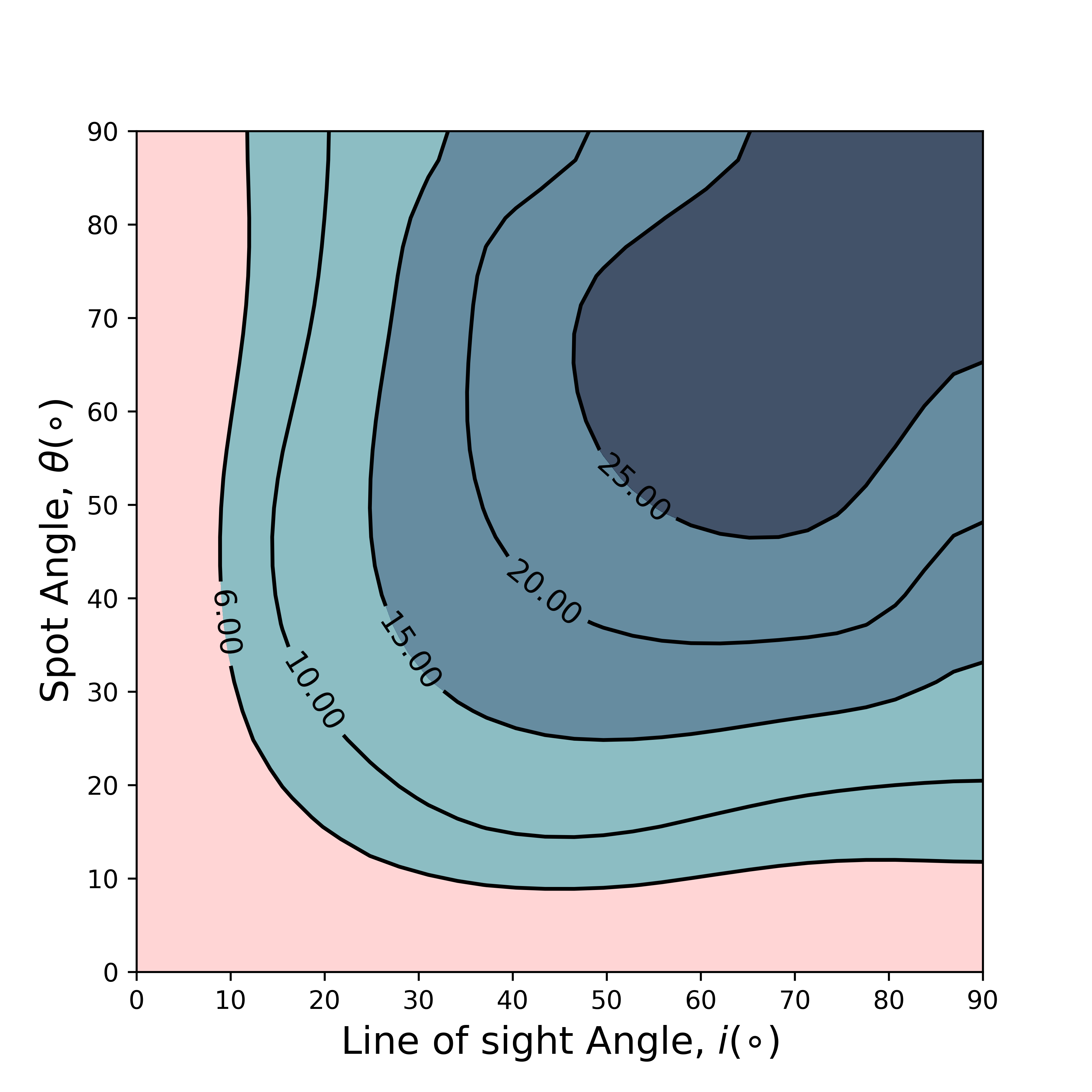}\quad
}
\subcaptionbox{J1601}{
   \includegraphics[width=0.3\linewidth]{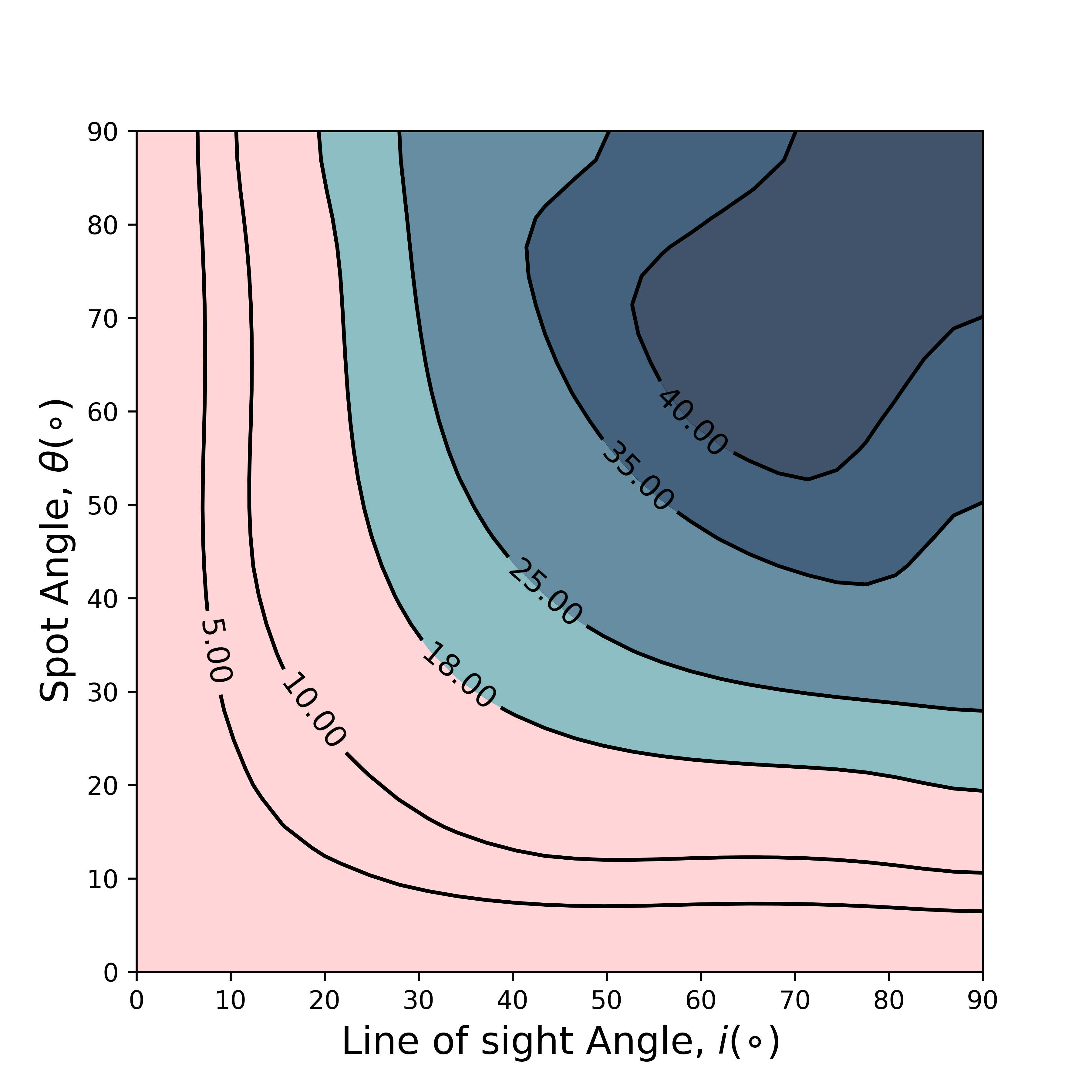}\quad
}
\subcaptionbox{J1713}{
	\includegraphics[width=0.3\linewidth]{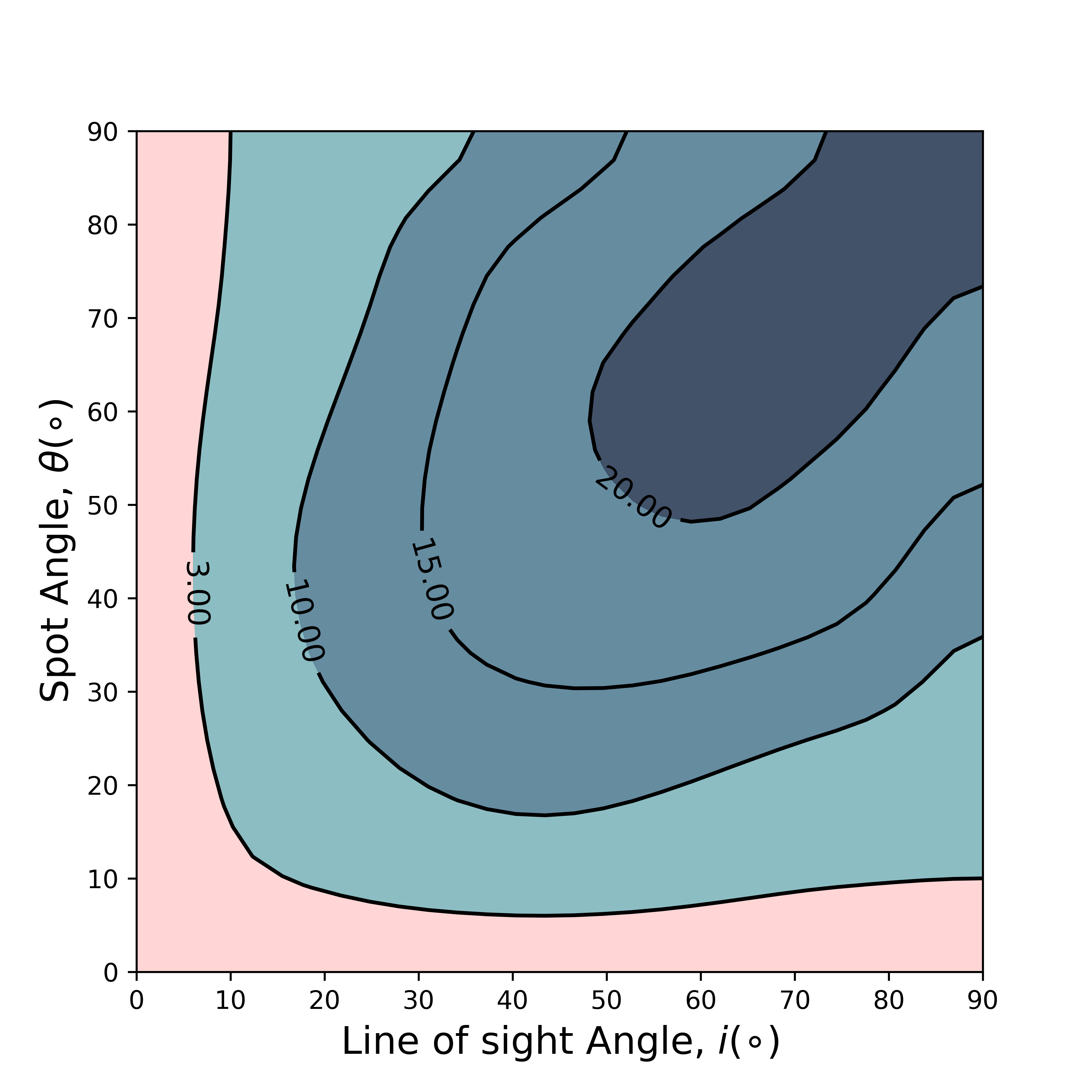}\quad
}
\smallskip
\subcaptionbox{J1732}{
   \includegraphics[width=0.3\linewidth]{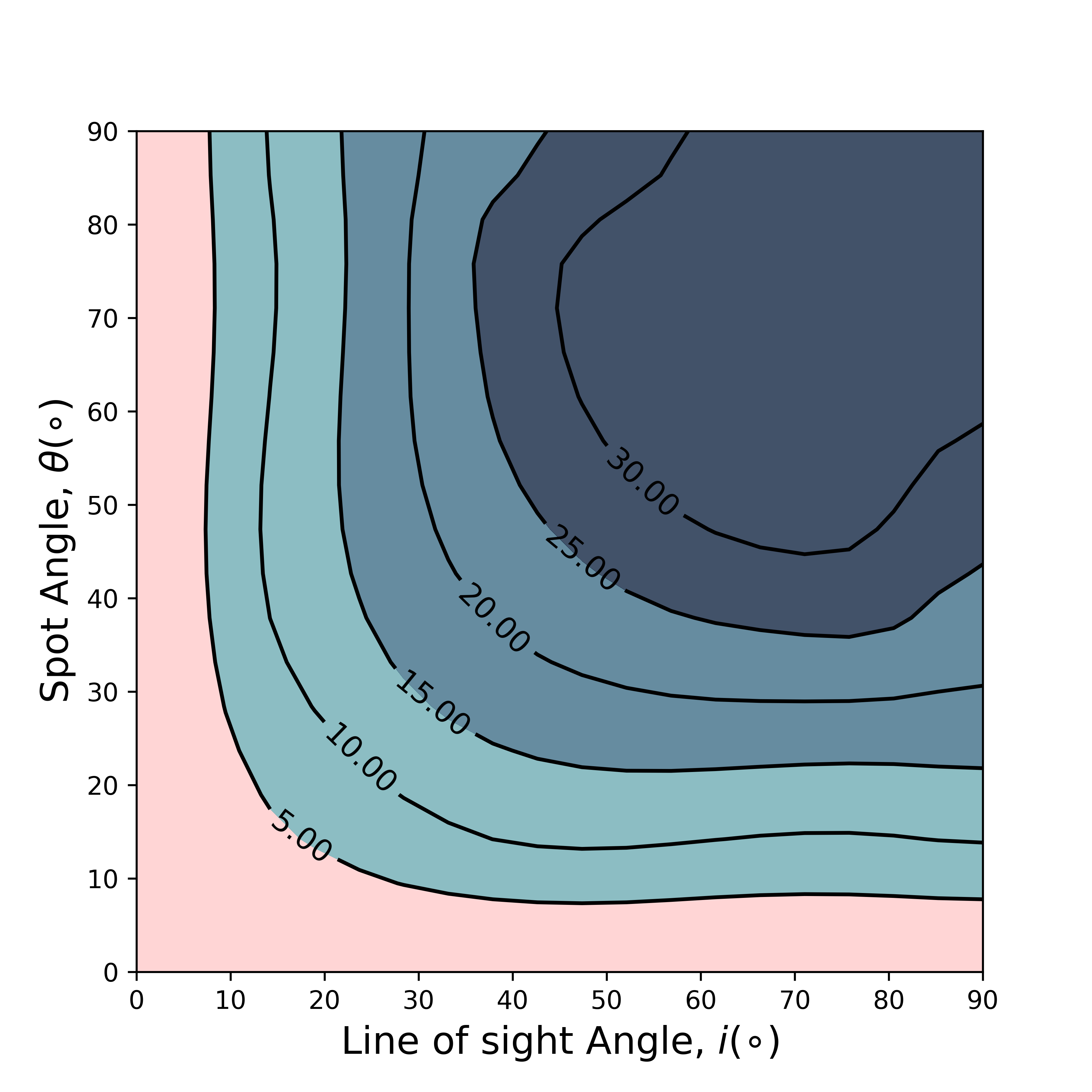}\quad
}
\subcaptionbox{J2323}{
	\includegraphics[width=0.3\linewidth]{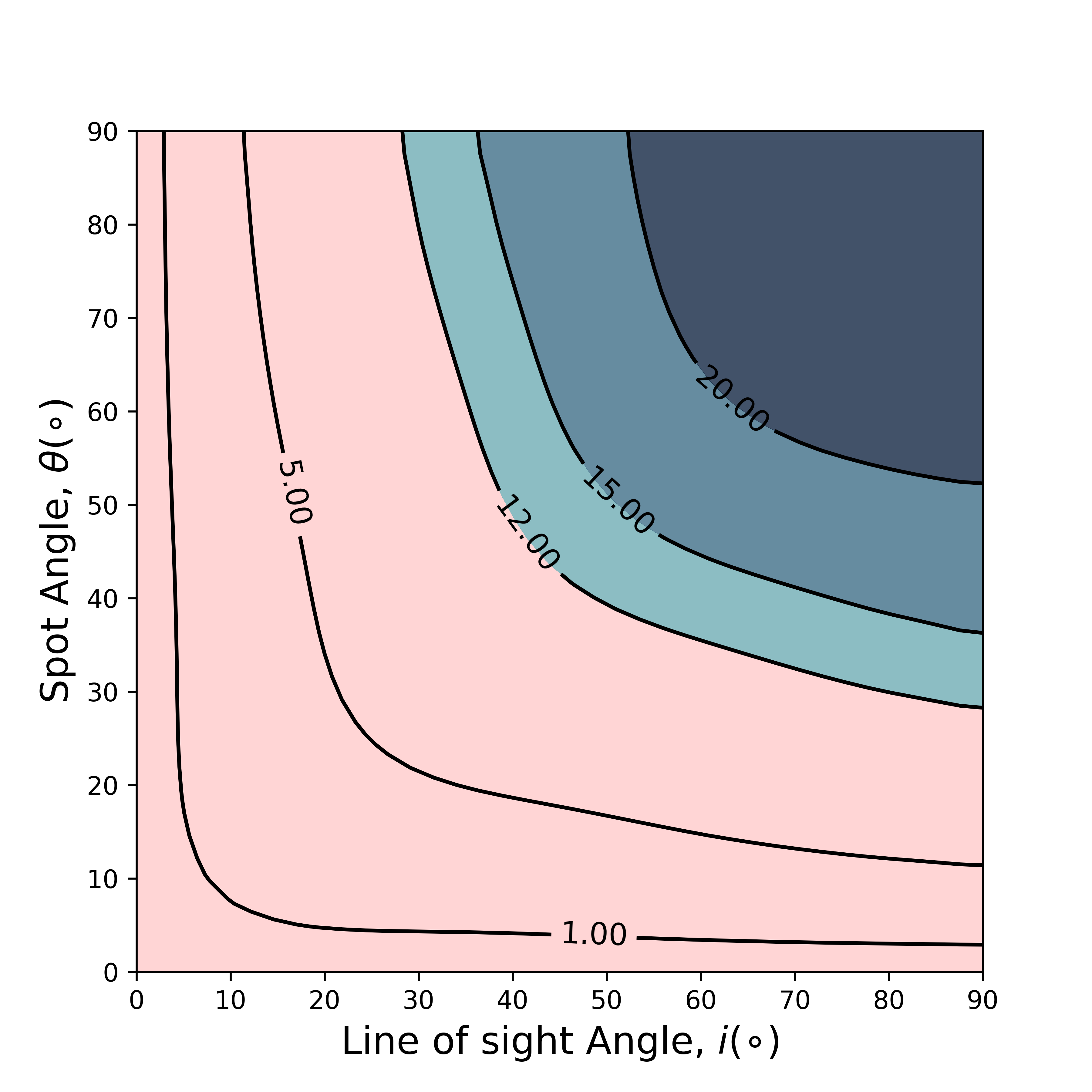}\quad
}
\caption{Contours of constant pulsed fraction in the $(i,\theta_{\mathrm{B}})$ plane assuming the surface inhomogeneous temperature distributions are caused by the ``crustal'' magnetic field. The pulsed fraction labels are given in \%. The allowed regions within the pulsed fraction limits are shown in the pink region. }
\label{fig:pfmap_mod2}
\end{figure}

 The results are presented in Fig~\ref{fig:pfmap_3bb} and \ref{fig:pfmap_mod2}. We plot the obtained $PF$ in the (i, $\theta_{B}$) plane and indicate the equal $PF$ with contours. The regions permitted by the observations are shown in pink.

\par By integrating all possible random orientations of line-of-sight inclination and spot angles where the calculated pulsed fraction is smaller than the upper limit from the timing analysis, we obtain the probability that the absence of pulsed signals is due to an unfavorable viewing geometry (e.g., \citealt{suleimanov2017AA43S}). 
For the `ad-hoc' and `crustal' models, we present the calculated probabilities in Table~\ref{tab:geo_prob}, and the probabilities are given in \%. As discussed by \cite{doroshenko2018AA76D} and \cite{suleimanov2017AA43S}, the probabilities of unfavorable viewing geometry are nonnegligible but relatively small for considering all sources, such as the CCO in SNRs Cas A (J2323), HESS J1731-347 (J1732), G330.2+1 (J1601). Taking these three CCOs in consideration, the probability is about $0.3\%$ (\citealt{doroshenko2018AA76D}). While we cannot exclude the presence of hot spots individually for sources with poorly constrained pulsed fractions (e.g., $\mathrm{PF}<3\%$ for J1713, and $5\%$ for J1732, see Table~\ref{tab:Timing_results}). As \cite{doroshenko2018AA76D} and \cite{suleimanov2017AA43S} did, we can compute the joint probability that the non-detection of pulsations in all five CCOs is due to unfavorable viewing geometries, assuming that the multi-temperature model is correct. Considering the results for both the `ad-hoc' and `crustal' models (Table~\ref{tab:geo_prob}), this number is quite small, less than $10^{-6}$. This conclusion seems robust against the multi-temperature emission model (see, e.g., \citealt{doroshenko2018AA76D}).

\begin{table}[htb]
    \centering\small
    \caption{Probability for each target in `ad-hoc' and `crustal' model}
    \label{tab:geo_prob}
    \begin{tabular}{c|ccccc|c}
    \hline
                     & J0852 & J1601 & J1713 & J1732 & J2323   \\ \hline
      `ad-hoc' (\%)  & 1.7   &   90  &  11.2 & 0.4   & 18.2  & \multirow{2}{*}{$<10^{-4}$} \\
      `crustal' (\%) & 4.24  &  14.8 &   1   & 0.4   & 28.4  & \\
      
    \hline
    \end{tabular}
    \note{The probabilities for each object are given in \%. And the joint probability is shown in the last column in \%. }
\end{table}

\subsection{The carbon atmosphere}
\par The inferred emission radii in the investigated targets (see Table~\ref{tab:bbodyrad} in Section~\ref{sect:spec}) are within 0.1 to 1.5 km for a two-blackbody model, smaller than the canonical radius of a neutron star ($10-15\,$km, \citealt{ozel2016ARAA..54..401O}). Two hot spots with different sizes and temperatures could be responsible for most of the measured flux in these sources. In this case, a third blackbody component ($95-138\,$eV, except for J1720 and J1818) representing the colder surface can be added to the model, which may be absorbed by the high hydrogen column density towards the targets ($n_{\mathrm{H}}=0.5-5.0\times10^{22}\,\mathrm{cm^{-2}}$). We verified in Section~\ref{sect:geo} that such a description would give rise to significant pulsed fractions ($20\%-50\%$), which, for most cases, could only be missed by the current generation of X-ray observatories for a few possible viewing geometries (Fig.~\ref{fig:pfmap_3bb} and ~\ref{fig:pfmap_mod2}, Table~\ref{tab:geo_prob}).

\par If the absence of pulsations in most CCOs cannnot be simply attributed to an unfavorable viewing geometry, we shall favour an instrinsically homogeneous temperature distribution on the neutron star surface. As discussed by \cite{Ho2009Natur71H}, for a neutron star with a carbon atmosphere, the X-ray emission could originate from the entire surface as the star rotates, despite local anisotropies.  

\par Such a model has successfully described the spectrum of the CCOs J1732, J1601 and J1818 (see also \citealt{klochkov2015AA53K,doroshenko2018AA76D,klochkov2016AA12K}). In this work, we also applied it to J1720, J1713, and J0852. We found the carbon atmosphere model characterize their spectra well, with $\chi_{\nu}^{2}\sim0.9-1.3$. Moreover, the obtained temperature, $145-200\,$eV, is more acceptable than the single or double component blackbody model for such a young neutron star, assuming standard cooling model ($<90-110\,$eV, \citealt{yakovlev2004ARAA169Y}), even though the derived distances of J1713 and J0852 are a bit higher than expected of the associated SNR by a factor of $2-3$. Nevertheless, only the carbon atmosphere model yields an acceptable distance for J1818, as discussed by \cite{klochkov2016AA12K}.

\par The presence of a carbon atmosphere could be understood within the context of hypercritical fall-back accretion and the ``hidden'' magnetic field model for CCOs (see, e.g., \citealt{geppert2004AA267G,Chevalier1993ApJ33C}). 
Lacking intense magnetospheric activity and pulsar wind nebulae suggest a weak magnetic field on the surface (\citealt{Ho2009Natur71H}). However, the strong magnetic field buried underneath the deep surface could be possible but has not yet emerged (\citealt{muslimov1995ApJ77M}). The hypercritical accretion would submerge the magnetic field soon after the supernova explosion, and the magnetic field is expected to diffuse back to the surface as the accretion weakens (\citealt{chevalier1989ApJ...346..847C,muslimov1995ApJ77M}). This re-diffusion process would happen on the time-scale of $\mathrm{10^3-10^4}$ years (\citealt{geppert1999AA}). For comparison, those CCOs without detected periods have an age range from 0.3 -- 6 kyr (e.g., J1713 with an age of $\sim1.6-2.1\,$kyr, \citealt{Tsuji2016PASJ108T}; J1720 with an age $<0.6\,$kyr, \citealt{Borkowski2020ApJ19B}), while the ages of CCOs with known periods are range from 4.5 kyr (\citealt{Becker2012ApJ141B}) to 7 kyr with an uncertainty factor of 3 (\citealt{Roger1988ApJ940R}). By comparing their ages, the CCOs with no periods detected may be undergoing re-diffusion progress of the magnetic field.

\par \cite{Ho2011MNRAS2567H} noted that, for three CCOs with measured spin periods, the bolometric luminosity is higher than its spin-down power of a factor of 3--5, even up to thousands. It could be understood as it is possible to submerge a higher magnetic field underneath the surface by accretion so that to explain the slow period derivatives. On the other hand, accretion is a way to support the relatively higher luminosity if the periodic characteristics could be extend to those CCOs with no periods detected. As discussed by \cite{doroshenko2016MNRAS2565D}, the brightest known CCO J1732 (up to $\mathrm{10^{34}\ erg\ s^{-1}}$) could be indirectly explained by powerful accretion when it was born in a binary system. In addition, assuming the gravitational energy of the inflowing matter (i.e., $GM\dot{M}/R_{ns}$) is responsible for the luminosity, it implies that an accretion rate larger than $10^{-13}\ \mathrm{M_{\odot}\,yr^{-1}}$ could account for the relatively steady X-ray emission of a magnitude of $\mathrm{10^{33}\ erg\ s^{-1}}$. This rate could be expected to reach $10^{-3}\ \mathrm{M_{\odot}\ yr^{-1}}$ in a common envelope (\citealt{Chevalier1993ApJ33C}), or $10^{-4}\ \mathrm{M_{\odot}\ yr^{-1}}$ in the early stage of supernova explosion (\citealt{chevalier1989ApJ...346..847C}). Therefore, accretion could be a possible way to explain the slow spin-down characteristic and the corresponding bolometric luminosity.

\par Moreover, a higher column density obtained for the CCOs applying with the carbon atmosphere model, comparing with the extended emission around the CCOs, may imply an additional absorption component in the line of sight, which has been discussed by \cite{doroshenko2018AA76D} for the CCOs J1601 and J1732. An additional absorption, even up to 20\%–-30\%, may be due to the existence of, e.g., inflow materials or surrounding remnants. An interaction is expected for the CCOs and the surrounding dust shell so that additional enhanced heating of the dust could be observed in the infrared band (\citealt{doroshenko2016MNRAS2565D}). Even though there is no direct evidence for the existence of surrounding material, an additional component seems like a natural way to explain the relatively higher column density in atmosphere models.

\par Finally, \cite{Ho2021MNRAS5015H} also noted that, for the CCO J1713, the slowly increasing magnetic field could be the reason for the gradual increase in temperature. Meanwhile, gravitational wave searches detected an outlier signal of J1713, which may imply a surface magnetic field of $\sim6\times10^{11}\,$G (\citealt{papa2020ApJ22P}), though it is higher than three detected CCOs. The magnetic field evolution of the CCOs can be understood with the above scenario, where a strong magnetic field is submerged underneath the surface soon after being born and then back to it with a timescale of $10^{3}$--$10^4$ years (e.g., \citealt{Ho2011MNRAS2567H,geppert1999AA,geppert2004AA267G,Pons2009AA207P,perna2013MNRAS2362P}).

\section{Conclusions}
\label{sect:conclusion}
\par We have conducted a systematic analysis on existing XMM-Newton, Chandra, and NICER observations of seven central compact objects in supernova remnants to investigate the reasons behind their lack of detected pulsations. Compared to previous works, we improved limits on the pulsed fractions of 1WGA J1713.4-3949, XMMU J173203.3-344518, and XMMU J172054.5-372652.  
We used a multi-component blackbody and atmosphere model to fit the spectra of seven CCOs, and it is hard to determine which one is better based on similar $\chi_{\nu}^2$. A third blackbody component is necessary to make it more natural and reasonable, even though the flux contribution of the third component seems negligible. Meanwhile, whether it is the cooler component of the 3-bb model or the carbon atmosphere model, the temperature obtained is consistent with that derived when we assume the standard cooling of young neutron stars. Furthermore, the updated limits and spectral properties of the sources were used to investigate the viability of multi-temperature emission models and the presence of hot spots. 
The possibility that all five CCOs (except for J1818 and J1720) have unfavorable viewing geometry is relatively tiny (less than $10^{-6}$), which seems impossible. While missing pulsations for individual CCOs in this group could be explained by this process, it is unlikely for all of these CCOs. 

\par Alternatively, the carbon atmosphere model can describe the emission spectra of CCOs and derive a more reasonable emission radius than the blackbody model. The obtained temperature also approaches the value given by the standard cooling curve for young neutron stars. Compared to the inferred emission radius of the two-blackbody model ($0.1-1.5\,$km), the emission size derived from the carbon atmosphere model close to a classical neutron star. While notably, the distance derived for 1WGA J1713.4-3949 and CXOU J085201.4-461753 mismatch with previous work and is a factor of $2-3$ higher than the associated supernova remnants. While there is no direct evidence of the heavier elements on the surface, the model provides more possibility to explain their spectra. Moreover, the occurrence of carbon on the surface could result from fallback accretion after the supernova explosion, which needs more observations to investigate.

\par No matter in what scenarios, the absence of pulsations can not simply be attributed to unfavorable viewing geometry or homogeneous temperature distribution. Both of them may contribute to the absence of pulsation jointly. The fallback accretion, however, could also lead to the inhomogeneous temperature distribution on the surface and further pulsation as the neutron star rotates. In contrast, the unfavorable viewing angles would reduce the observable pulsed fraction. While in this case, further searches are necessary to constraint the upper limits and even the composition on the surface.

\normalem
\begin{acknowledgements}
We thank the anonymous referee for useful comments and suggestions, which helped to improve the paper. AMP gratefully acknowledges support from the Chinese Academy of Science's President International Fellowship Initiative (CAS PIFI 2019VMC0008). SY thanks the National Natural Science Foundation of China for financial support (grants U1838112). QW appreciates the use of computer facilities at Purple Mountain Observatory.
\end{acknowledgements}

\appendix

\bibliographystyle{raa}
\bibliography{ms2021-0223R1}

\begin{thebibliography}{82}
\providecommand\natexlab[1]{#1}
\providecommand\JournalTitle[1]{#1}

\bibitem[{Acero} {et~al.}(2013)]{Acero2013AA7A}
{Acero}, F., {Gallant}, Y., {Ballet}, J., {Renaud}, M., \& {Terrier}, R. 2013,
  \aap, 551, A7

\bibitem[{Acero} {et~al.}(2015)]{Acero2015AA74A}
{Acero}, F., {Lemoine-Goumard}, M., {Renaud}, M., {et~al.} 2015, \aap, 580, A74

\bibitem[{Allen} {et~al.}(2015)]{Allen2015ApJ82A}
{Allen}, G.~E., {Chow}, K., {DeLaney}, T., {et~al.} 2015, \apj, 798, 82

\bibitem[{Arnaud}(1996)]{Arnaud1996ASPC17A}
{Arnaud}, K.~A. 1996, in Astronomical Society of the Pacific Conference Series,
  Vol. 101, Astronomical Data Analysis Software and Systems V, ed. G.~H.
  {Jacoby} \& J.~{Barnes}, 17

\bibitem[{Ashworth}(1980)]{Ashworth1980JHA1A}
{Ashworth}, W.~B., J. 1980, Journal for the History of Astronomy, 11, 1

\bibitem[{Becker} {et~al.}(2012)]{Becker2012ApJ141B}
{Becker}, W., {Prinz}, T., {Winkler}, P.~F., \& {Petre}, R. 2012, \apj, 755,
  141

\bibitem[{Beloborodov}(2002)]{Beloborodov2002ApJ85B}
{Beloborodov}, A.~M. 2002, \apjl, 566, L85

\bibitem[{Bogdanov}(2014)]{Bogdanov2014ApJ94B}
{Bogdanov}, S. 2014, \apj, 790, 94

\bibitem[{Borkowski} {et~al.}(2020)]{Borkowski2020ApJ19B}
{Borkowski}, K.~J., {Miltich}, W., \& {Reynolds}, S.~P. 2020, \apjl, 905, L19

\bibitem[{Borkowski} {et~al.}(2018)]{Borkowski2018ApJ21B}
{Borkowski}, K.~J., {Reynolds}, S.~P., {Williams}, B.~J., \& {Petre}, R. 2018,
  \apjl, 868, L21

\bibitem[{Buccheri} {et~al.}(1983)]{Buccheri1983AA}
{Buccheri}, R., {Bennett}, K., {Bignami}, G.~F., {et~al.} 1983, \aap, 128, 245

\bibitem[{Cassam-Chena{\"\i}} {et~al.}(2004)]{cassam2004aAA}
{Cassam-Chena{\"\i}}, G., {Decourchelle}, A., {Ballet}, J., {et~al.} 2004,
  \aap, 427, 199

\bibitem[{Caswell} {et~al.}(1982)]{Caswell1982MNRAS1143C}
{Caswell}, J.~L., {Haynes}, R.~F., {Milne}, D.~K., \& {Wellington}, K.~J. 1982,
  \mnras, 200, 1143

\bibitem[{Chevalier}(1989)]{chevalier1989ApJ...346..847C}
{Chevalier}, R.~A. 1989, \apj, 346, 847

\bibitem[{Chevalier}(1993)]{Chevalier1993ApJ33C}
{Chevalier}, R.~A. 1993, \apjl, 411, L33

\bibitem[{de Luca}(2008)]{deluca2008AIPC..983..311D}
{de Luca}, A. 2008, in American Institute of Physics Conference Series, Vol.
  983, 40 Years of Pulsars: Millisecond Pulsars, Magnetars and More, ed.
  C.~{Bassa}, Z.~{Wang}, A.~{Cumming}, \& V.~M. {Kaspi}, 311

\bibitem[{De Luca}(2017)]{deluca2017JPhCS}
{De Luca}, A. 2017, in Journal of Physics Conference Series, Vol. 932, Journal
  of Physics Conference Series, 012006

\bibitem[{Doroshenko} {et~al.}(2016)]{doroshenko2016MNRAS2565D}
{Doroshenko}, V., {P{\"u}hlhofer}, G., {Kavanagh}, P., {et~al.} 2016, \mnras,
  458, 2565

\bibitem[{Doroshenko} {et~al.}(2018)]{doroshenko2018AA76D}
{Doroshenko}, V., {Suleimanov}, V., \& {Santangelo}, A. 2018, \aap, 618, A76

\bibitem[{Elshamouty} {et~al.}(2016)]{Elshamouty2016ApJ162E}
{Elshamouty}, K.~G., {Heinke}, C.~O., {Morsink}, S.~M., {Bogdanov}, S., \&
  {Stevens}, A.~L. 2016, \apj, 826, 162

\bibitem[{Fesen} {et~al.}(2006)]{fesen2006ApJ...636..848F}
{Fesen}, R.~A., {Pavlov}, G.~G., \& {Sanwal}, D. 2006, \apj, 636, 848

\bibitem[{Gaensler} {et~al.}(2008)]{gaensler2008ApJ37G}
{Gaensler}, B.~M., {Tanna}, A., {Slane}, P.~O., {et~al.} 2008, \apjl, 680, L37

\bibitem[{Garmire} {et~al.}(2003)]{Garmire2003SPIE28G}
{Garmire}, G.~P., {Bautz}, M.~W., {Ford}, P.~G., {Nousek}, J.~A., \& {Ricker},
  George~R., J. 2003, in Society of Photo-Optical Instrumentation Engineers
  (SPIE) Conference Series, Vol. 4851, X-Ray and Gamma-Ray Telescopes and
  Instruments for Astronomy., ed. J.~E. {Truemper} \& H.~D. {Tananbaum}, 28

\bibitem[{Gendreau} \& {Arzoumanian}(2017)]{Gendreau2017NatAs895G}
{Gendreau}, K., \& {Arzoumanian}, Z. 2017, Nature Astronomy, 1, 895

\bibitem[{Geppert} {et~al.}(2004)]{geppert2004AA267G}
{Geppert}, U., {K{\"u}ker}, M., \& {Page}, D. 2004, \aap, 426, 267

\bibitem[{Geppert} {et~al.}(1999)]{geppert1999AA}
{Geppert}, U., {Page}, D., \& {Zannias}, T. 1999, \aap, 345, 847

\bibitem[{Gotthelf} \& {Halpern}(2009)]{gotthelf2009ApJ...695L..35G}
{Gotthelf}, E.~V., \& {Halpern}, J.~P. 2009, \apjl, 695, L35

\bibitem[{Gotthelf} {et~al.}(2013)]{Gotthelf2013ApJ58G}
{Gotthelf}, E.~V., {Halpern}, J.~P., \& {Alford}, J. 2013, \apj, 765, 58

\bibitem[{Gotthelf} {et~al.}(2005)]{gotthelf2005ApJ390G}
{Gotthelf}, E.~V., {Halpern}, J.~P., \& {Seward}, F.~D. 2005, \apj, 627, 390

\bibitem[{Greenstein} \& {Hartke}(1983)]{Greenstein1983ApJ...271..283G}
{Greenstein}, G., \& {Hartke}, G.~J. 1983, \apj, 271, 283

\bibitem[{Halpern} \& {Gotthelf}(2010{\natexlab{a}})]{Halpern2010ApJ436H}
{Halpern}, J.~P., \& {Gotthelf}, E.~V. 2010{\natexlab{a}}, \apj, 709, 436

\bibitem[{Halpern} \& {Gotthelf}(2010{\natexlab{b}})]{halpern2010ApJ941H}
{Halpern}, J.~P., \& {Gotthelf}, E.~V. 2010{\natexlab{b}}, \apj, 710, 941

\bibitem[{Halpern} \& {Gotthelf}(2015)]{halpern2015ApJ61H}
{Halpern}, J.~P., \& {Gotthelf}, E.~V. 2015, \apj, 812, 61

\bibitem[{Heinke} \& {Ho}(2010)]{heinke2010ApJ167H}
{Heinke}, C.~O., \& {Ho}, W. C.~G. 2010, \apjl, 719, L167

\bibitem[{Ho}(2011)]{Ho2011MNRAS2567H}
{Ho}, W. C.~G. 2011, \mnras, 414, 2567

\bibitem[{Ho} \& {Heinke}(2009)]{Ho2009Natur71H}
{Ho}, W. C.~G., \& {Heinke}, C.~O. 2009, \nat, 462, 71

\bibitem[{Ho} {et~al.}(2021)]{Ho2021MNRAS5015H}
{Ho}, W. C.~G., {Zhao}, Y., {Heinke}, C.~O., {et~al.} 2021, \mnras, 506, 5015

\bibitem[{Jansen} {et~al.}(2001)]{Jansen2001AA1J}
{Jansen}, F., {Lumb}, D., {Altieri}, B., {et~al.} 2001, \aap, 365, L1

\bibitem[{Jethwa} {et~al.}(2015)]{Jethwa2015AA104J}
{Jethwa}, P., {Saxton}, R., {Guainazzi}, M., {Rodriguez-Pascual}, P., \&
  {Stuhlinger}, M. 2015, \aap, 581, A104

\bibitem[{Kargaltsev} {et~al.}(2002)]{kargaltsev2002ApJ}
{Kargaltsev}, O., {Pavlov}, G.~G., {Sanwal}, D., \& {Garmire}, G.~P. 2002,
  \apj, 580, 1060

\bibitem[{Klochkov} {et~al.}(2013)]{klochkov2013AA}
{Klochkov}, D., {P{\"u}hlhofer}, G., {Suleimanov}, V., {et~al.} 2013, \aap,
  556, A41

\bibitem[{Klochkov} {et~al.}(2015)]{klochkov2015AA53K}
{Klochkov}, D., {Suleimanov}, V., {P{\"u}hlhofer}, G., {et~al.} 2015, \aap,
  573, A53

\bibitem[{Klochkov} {et~al.}(2016)]{klochkov2016AA12K}
{Klochkov}, D., {Suleimanov}, V., {Sasaki}, M., \& {Santangelo}, A. 2016, \aap,
  592, L12

\bibitem[{Lazendic} {et~al.}(2003)]{lazendic2003ApJ}
{Lazendic}, J.~S., {Slane}, P.~O., {Gaensler}, B.~M., {et~al.} 2003, \apjl,
  593, L27

\bibitem[{Lovchinsky} {et~al.}(2011)]{lovchinsky2011ApJ70L}
{Lovchinsky}, I., {Slane}, P., {Gaensler}, B.~M., {et~al.} 2011, \apj, 731, 70

\bibitem[{Maxted} {et~al.}(2018)]{Maxted2018MNRAS662M}
{Maxted}, N., {Burton}, M., {Braiding}, C., {et~al.} 2018, \mnras, 474, 662

\bibitem[{McClure-Griffiths} {et~al.}(2001)]{McClure2001ApJ394M}
{McClure-Griffiths}, N.~M., {Green}, A.~J., {Dickey}, J.~M., {et~al.} 2001,
  \apj, 551, 394

\bibitem[{Muslimov} \& {Page}(1995)]{muslimov1995ApJ77M}
{Muslimov}, A., \& {Page}, D. 1995, \apjl, 440, L77

\bibitem[{Ostriker} \& {Gunn}(1969)]{ostriker1969ApJ1395}
{Ostriker}, J.~P., \& {Gunn}, J.~E. 1969, \apj, 157, 1395

\bibitem[{{\"O}zel} \& {Freire}(2016)]{ozel2016ARAA..54..401O}
{{\"O}zel}, F., \& {Freire}, P. 2016, \araa, 54, 401

\bibitem[{Papa} {et~al.}(2020)]{papa2020ApJ22P}
{Papa}, M.~A., {Ming}, J., {Gotthelf}, E.~V., {et~al.} 2020, \apj, 897, 22

\bibitem[{Park} {et~al.}(2009)]{park2009ApJ431P}
{Park}, S., {Kargaltsev}, O., {Pavlov}, G.~G., {et~al.} 2009, \apj, 695, 431

\bibitem[{Pavlov} {et~al.}(2001)]{Pavlov2001ApJ131P}
{Pavlov}, G.~G., {Sanwal}, D., {K{\i}z{\i}ltan}, B., \& {Garmire}, G.~P. 2001,
  \apjl, 559, L131

\bibitem[{Pavlov} {et~al.}(2000)]{Pavlov2000ApJ53P}
{Pavlov}, G.~G., {Zavlin}, V.~E., {Aschenbach}, B., {Tr{\"u}mper}, J., \&
  {Sanwal}, D. 2000, \apjl, 531, L53

\bibitem[{Pavlov} {et~al.}(1999)]{Pavlov1999ApJ45P}
{Pavlov}, G.~G., {Zavlin}, V.~E., \& {Tr{\"u}mper}, J. 1999, \apjl, 511, L45

\bibitem[{P{\'e}rez-Azor{\'\i}n} {et~al.}(2006)]{Perez2006AA1009P}
{P{\'e}rez-Azor{\'\i}n}, J.~F., {Miralles}, J.~A., \& {Pons}, J.~A. 2006, \aap,
  451, 1009

\bibitem[{Perna} {et~al.}(2013)]{perna2013MNRAS2362P}
{Perna}, R., {Vigan{\`o}}, D., {Pons}, J.~A., \& {Rea}, N. 2013, \mnras, 434,
  2362

\bibitem[{Pires} {et~al.}(2019)]{Pires2019AA73P}
{Pires}, A.~M., {Schwope}, A.~D., {Haberl}, F., {et~al.} 2019, \aap, 623, A73

\bibitem[{Pons} {et~al.}(2009)]{Pons2009AA207P}
{Pons}, J.~A., {Miralles}, J.~A., \& {Geppert}, U. 2009, \aap, 496, 207

\bibitem[{Posselt} \& {Pavlov}(2018)]{Posselt2018ApJ135P}
{Posselt}, B., \& {Pavlov}, G.~G. 2018, \apj, 864, 135

\bibitem[{Posselt} {et~al.}(2013)]{Posselt2013ApJ186P}
{Posselt}, B., {Pavlov}, G.~G., {Suleimanov}, V., \& {Kargaltsev}, O. 2013,
  \apj, 779, 186

\bibitem[{Potekhin} {et~al.}(2020)]{Potekhin2020MNRAS5052P}
{Potekhin}, A.~Y., {Zyuzin}, D.~A., {Yakovlev}, D.~G., {Beznogov}, M.~V., \&
  {Shibanov}, Y.~A. 2020, \mnras, 496, 5052

\bibitem[{Reed} {et~al.}(1995)]{Reed1995ApJ706R}
{Reed}, J.~E., {Hester}, J.~J., {Fabian}, A.~C., \& {Winkler}, P.~F. 1995,
  \apj, 440, 706

\bibitem[{Reynolds} {et~al.}(2006)]{reynolds2006ApJ45R}
{Reynolds}, S.~P., {Borkowski}, K.~J., {Hwang}, U., {et~al.} 2006, \apjl, 652,
  L45

\bibitem[{Roger} {et~al.}(1988)]{Roger1988ApJ940R}
{Roger}, R.~S., {Milne}, D.~K., {Kesteven}, M.~J., {Wellington}, K.~J., \&
  {Haynes}, R.~F. 1988, \apj, 332, 940

\bibitem[{Sasaki} {et~al.}(2018)]{Sasaki2018MNRAS3033S}
{Sasaki}, M., {M{\"a}kel{\"a}}, M.~M., {Klochkov}, D., {Santangelo}, A., \&
  {Suleimanov}, V. 2018, \mnras, 479, 3033

\bibitem[{Schwope} {et~al.}(2005)]{Schwope2005AA597S}
{Schwope}, A.~D., {Hambaryan}, V., {Haberl}, F., \& {Motch}, C. 2005, \aap,
  441, 597

\bibitem[{Slane} {et~al.}(1999)]{Slane1999ApJ357S}
{Slane}, P., {Gaensler}, B.~M., {Dame}, T.~M., {et~al.} 1999, \apj, 525, 357

\bibitem[{Slane} {et~al.}(2001)]{Slane2001ApJ814S}
{Slane}, P., {Hughes}, J.~P., {Edgar}, R.~J., {et~al.} 2001, \apj, 548, 814

\bibitem[{Str{\"u}der} {et~al.}(2001)]{Struder2001AA18S}
{Str{\"u}der}, L., {Briel}, U., {Dennerl}, K., {et~al.} 2001, \aap, 365, L18

\bibitem[{Suleimanov} {et~al.}(2010)]{Suleimanov2010AA}
{Suleimanov}, V.~F., {Hambaryan}, V., {Potekhin}, A.~Y., \& {Werner}, K. 2010,
  \aap, 522, A111

\bibitem[{Suleimanov} {et~al.}(2014)]{suleimanov2014ApJS13S}
{Suleimanov}, V.~F., {Klochkov}, D., {Pavlov}, G.~G., \& {Werner}, K. 2014,
  \apjs, 210, 13

\bibitem[{Suleimanov} {et~al.}(2017)]{suleimanov2017AA43S}
{Suleimanov}, V.~F., {Klochkov}, D., {Poutanen}, J., \& {Werner}, K. 2017,
  \aap, 600, A43

\bibitem[{Sun} {et~al.}(2004)]{Sun2004ApJ742S}
{Sun}, M., {Seward}, F.~D., {Smith}, R.~K., \& {Slane}, P.~O. 2004, \apj, 605,
  742

\bibitem[{Tsuji} \& {Uchiyama}(2016)]{Tsuji2016PASJ108T}
{Tsuji}, N., \& {Uchiyama}, Y. 2016, \pasj, 68, 108

\bibitem[{Wang} {et~al.}(1997)]{wangZR1997AA}
{Wang}, Z.~R., {Qu}, Q.~Y., \& {Chen}, Y. 1997, \aap, 318, L59

\bibitem[{Weisskopf} {et~al.}(2002)]{Weisskopf2002PASP1W}
{Weisskopf}, M.~C., {Brinkman}, B., {Canizares}, C., {et~al.} 2002, \pasp, 114,
  1

\bibitem[{Wijngaarden} {et~al.}(2019)]{Wijngaarden2019MNRAS974W}
{Wijngaarden}, M.~J.~P., {Ho}, W. C.~G., {Chang}, P., {et~al.} 2019, \mnras,
  484, 974

\bibitem[{Williams} {et~al.}(2018)]{Williams2018ApJ118W}
{Williams}, B.~J., {Hewitt}, J.~W., {Petre}, R., \& {Temim}, T. 2018, \apj,
  855, 118

\bibitem[{Wilms} {et~al.}(2000)]{Wilms2000ApJ914W}
{Wilms}, J., {Allen}, A., \& {McCray}, R. 2000, \apj, 542, 914

\bibitem[{Yakovlev} \& {Pethick}(2004)]{yakovlev2004ARAA169Y}
{Yakovlev}, D.~G., \& {Pethick}, C.~J. 2004, \araa, 42, 169

\bibitem[{Zavlin} {et~al.}(2000)]{Zavlin2000ApJ25Z}
{Zavlin}, V.~E., {Pavlov}, G.~G., {Sanwal}, D., \& {Tr{\"u}mper}, J. 2000,
  \apjl, 540, L25

\end{thebibliography}

\end{document}